\documentclass[final]{svjour2}
\usepackage{graphicx}
\usepackage{rotating}
\usepackage{amssymb}
\usepackage{amsmath}
\usepackage{epstopdf}
\usepackage{mathptmx}
\usepackage{hyperref}
\makeatletter
\journalname{Journal of Low Temperature Physics}
\begin{document}

\newcommand{\hdblarrow}{H\makebox[0.9ex][l]{$\downdownarrows$}-}
\title{Entrainment in Superfluid Neutron Star Crusts: Hydrodynamic Description and Microscopic Origin}

\author{N. Chamel}

\institute{Institut d'Astronomie et d'Astrophysique, CP-226, Universit\'e Libre de Bruxelles, \\
1050 Brussels, Belgium\\
\email{nchamel@ulb.ac.be}}

\maketitle

\begin{abstract}
In spite of the absence of viscous drag, the neutron superfluid permeating the inner crust of a neutron star
cannot flow freely, and is entrained by the nuclear lattice similarly to laboratory superfluid atomic gases in optical 
lattices. The role of entrainment on the neutron superfluid dynamics is reviewed. 
For this purpose, a minimal hydrodynamical model of superfluidity in neutron-star crusts is presented. 
This model relies on a fully four-dimensionally covariant action principle. The equivalence 
of this formulation with the more traditional approach is demonstrated. In addition, the different treatments of 
entrainment in terms of dynamical effective masses or superfluid density are clarified. 
The nuclear energy density functional theory employed for the calculations 
of all the necessary microscopic inputs is also reviewed, focusing on superfluid properties. In particular, 
the microscopic origin of entrainment and the different methods to estimate its importance are discussed. 
\end{abstract}

\section{Introduction}

Neutron stars, the compact stellar remnants of gravitational core collapse supernova explosions of massive stars 
(with a mass between 8 and and 10 times that of the Sun), contain matter under the most extreme conditions with 
central densities exceeding that prevailing in atomic nuclei~\cite{haensel2007}. 
The interior of a neutron star comprises essentially five distinct regions: (i) an ocean of liquid iron surmounted 
by a thin atmosphere of light elements; (ii) an outer crust, at densities ranging between $\sim 10^4$~g~cm$^{-3}$ and 
a few $10^{11}$~g~cm$^{-3}$, consisting of a dense plasma of neutron-rich nuclei arranged on a body-centered cubic 
lattice coexisting with a highly degenerate relativistic electron gas; (iii) an inner crust, composed of an inhomogeneous 
assembly of neutron-proton clusters immersed in a neutron liquid and neutralised by electrons; (iv) an outer core at 
densities above $\sim 10^{14}$ g.cm$^{-3}$ made of neutrons, with a small admixture of protons and leptons; (v) an 
inner core whose composition remains highly speculative. 

With typical temperatures of order $10^7$~K, the interior of a mature neutron star is expected to be cold enough for the 
existence of nuclear superfluid and superconducting phases (see, e.g. Ref.~\cite{chamel2017} for a recent review). In 
particular, the free neutrons in the inner crust are thought to become superfluid by forming Cooper pairs analogously to 
electrons in conventional superconductors. This prediction is supported by observations of giant pulsar frequency glitches, 
as in the emblematic Vela pulsar. Remarkably, similar sudden spin-ups have been observed in superfluid helium~\cite{tsakadze1980}. 
Glitches are usually interpreted as transfers of angular momentum between the neutron superfluid and the rest of star 
triggered by the unpinning of quantized vortices~\cite{anderson1975,alpar1985}. 
However, it has been recently realized that the neutron 
superfluid does not flow freely as previously assumed, but strongly interacts with the periodic nuclear lattice~\cite{chamel2005,chamel2006b,chamel2012}
similarly to superfluid cold atomic gases in optical lattices~\cite{watanabe2008}. Due to these entrainment effects, 
the neutron superfluid in the crust is not enough to explain giant pulsar glitches~\cite{andersson2012,chamel2013b,delsate2016}, 
suggesting that another superfluid reservoir in the stellar core is involved~\cite{gugercinoglu2014,ho2015,pizzochero2017}. 
The neutron superfluid may leave its imprint on other observed astrophysical phenomena such as the thermal relaxation of transiently 
accreting neutron stars during quiescence, or quasiperiodic oscillations in the hard X-ray emission detected in the tails of giant 
flares from a few soft-gamma ray repeaters (see, e.g. Ref.~\cite{chamel2017}). The interpretation of all these phenomena requires 
a better understanding of the dynamics of superfluid neutron stars. 

In this paper, recent developments in the understanding of entrainment effects in neutron star crusts are reviewed. The hydrodynamical aspects 
are discussed in Section~\ref{sec:hydro}. After briefly reviewing in Section~\ref{sec:hydro-principle} the convective variational 
action principle introduced by Brandon Carter~\cite{carter1989}, 
a minimal model of superfluid neutron-star crusts is presented in Section~\ref{sec:hydro-min-model}. The equivalence of this approach with 
the more heuristic formulation of Refs.~\cite{pethick2010,kobyakov2013} using the traditional space-time decomposition is demonstrated in 
Section~\ref{sec:hydro-equiv}. As an application, low-energy collective excitations are studied in Section~\ref{sec:hydro-excitations}. The 
calculations of all the necessary underlying microscopic inputs are discussed in Section~\ref{sec:micro}. In Section~\ref{sec:micro-EDF}, the 
 nuclear energy density functional (EDF) theory is reviewed. Its application to the description of neutron-star crusts and entrainment effects are 
discussed in Section~\ref{sec:micro-crust}. 

\section{Superfluid dynamics and entrainment in neutron-star crusts}
\label{sec:hydro}
\subsection{Convective variational principle}
\label{sec:hydro-principle}

The traditional approach to superfluid hydrodynamics blurring the distinction between velocity and momentum makes it difficult to adapt 
and extend Landau's original two-fluid model to the relativistic context, as required for a realistic description of neutron stars. 
For this purpose, Brandon Carter~\cite{carter1989} developed an elegant variational formalism based on exterior calculus
(see e.g. Refs.~\cite{carter2001,gourgoulhon2006,andersson2007} for a review). The action 
\begin{equation}\label{eq:action}
\mathcal{A}= \int \Lambda\{n_{_{\rm X}}^{\, \nu}\}\, {\rm d}\mathcal{M}^{(4)}\, ,
\end{equation}
is integrated over the 4-dimensional manifold $\mathcal{M}^{(4)}$, and the Lagrangian density $\Lambda$ (also referred to as the master function) 
depends on the 4-current vectors $n_{_{\rm X}}^{\, \nu}$ of the different fluids (with the Greek letter $\nu=0,1,2,3$ denoting the space-time components whereas 
the different constituents are labelled by X). The dynamical equations, 
obtained by requiring $\delta \mathcal{A}=0$ under infinitesimal variations of the fluid particle trajectories, take a very concise form
(summation over repeated indices will be assumed throughout this paper except for those labeling constituents): 
\begin{equation}\label{eq.hydro.covariant}
 n_{_{\rm X}}^{\, \mu}\varpi^{_{\rm X}}_{\!\mu\nu} + \pi^{_{\rm X}}_{\, \nu}\nabla_\mu n_{_{\rm X}}^{\, \mu}  = f^{_{\rm X}}_{\,\nu}\, , 
\end{equation}
expressed in terms of the total 4-momentum 1-form 
\begin{equation}\label{eq.full.momentum}
 \pi^{_{\rm X}}_{\, \mu} = \frac{\partial \Lambda}{\partial n_{_{\rm X}}^{\, \mu}}\, ,
\end{equation}
the vorticity 2-form 
\begin{equation}\label{eq.vorticity.2form}
 \varpi^{_{\rm X}}_{\!\mu\nu}= \nabla_{\!\mu}\pi^{_{\rm X}}_{\,\nu} - \nabla_{\!\nu}\pi^{_{\rm X}}_{\,\mu} \, ,
\end{equation}
and $f^{_{\rm X}}_{\,\nu}$ denotes the 4-force density 1-form acting on the fluids. 
As emphasized by Carter (see, e.g. Ref.~\cite{carter1994}), the fundamentally different physical natures of 
the velocity and the momentum are reflected in their mathematical structure: while the former belongs in a tangent bundle (vector), the latter 
belongs in a cotangent bundle (covector), as can be clearly seen from the definition (\ref{eq.full.momentum}). 

Carter's formalism was later adapted to the comparatively more intrincate Newtonian theory within a 4-dimensionally covariant framework~\cite{carter2004,carter2005}. 
This fully covariant description not only provides a direct comparison with the relativistic theory (see, e.g. Ref.~\cite{chamel2008}), but also 
 helped to reveal new conservation laws such as the conservation of generalised helicy currents in superfluid mixtures. Moreover, the derivation of various 
 identities (e.g. generalised Bernouilli constants and virial theorems) is considerably simplified by making use of mathematical concepts from differential geometry 
 that have been extremely fruitful in the relativistic context, such as Killing vectors (see, e.g., Ref.~\cite{chamel2015}). 
Although less accurate than a fully relativistic description, a Newtonian treatment of superfluid neutron stars can provide valuable insight at a much reduced computational cost. For this reason, 
studies of the neutron-star superfluid dynamics in Newtonian theory are still being carried out. 
The 4-current vector $n_{_{\rm X}}^{\, \mu}$ in Newtonian spacetime is obtained by combining the particle number 
density $n_{_{\rm X}}\equiv n_{_{\rm X}}^{\,0}$ with the 3-current vector $n_{_{\rm X}}^{\, i}=n_{_{\rm X}}v_{_{\rm X}}^{\, i}$ where $v_{_{\rm X}}^{\, i}$ is the 
corresponding velocity vector (with the Latin index $i=1,2,3$ indicating the spatial components). Because Carter's formalism relies on exterior calculus, the 
equations of motion (\ref{eq.hydro.covariant}) do not directly depend on the space-time metric, and thus take the same form in Newtonian theory. 
Dissipative processes (e.g. viscosity in non-superfluid constituents, superfluid vortex drag, mutual friction between non-superfluid constituents, nuclear 
reactions) can be treated within the same framework~\cite{carter2005b}. Carter's formalism was further extended in order to allow for the inclusion of the elasticity 
of the crust~\cite{carter2006b} and the presence of a strong magnetic field~\cite{carter2006}. The relativistic formalism was developed in Ref.~\cite{carter2006c}.

\subsection{Minimal model of superfluid neutron-star crusts}
\label{sec:hydro-min-model}

Although it would be necessary to account for gravity in a global description of neutron stars, its effects on the local superfluid dynamics of 
neutron-star crusts, on which we focus here, are relatively small and will thus be neglected. A smooth-averaged hydrodynamic treatment at length 
scales large compared with the mean ion spacing $a_{\rm I}=(3/(4 \pi n_{\rm I}))^{1/3}$ with $n_{\rm I}$ the ion number density, the neutron 
superfluid coherence length $\xi$, and the electron screening length $r_e=(4\pi e^2 dn_e/d\mu_e)^{-1/2}$ with $n_e$ the electron number density and 
$\mu_e$ the electron Fermi energy, was presented in Ref.~\cite{carter2006d}. 

In this model, 
the crust of a neutron star is described by two interpenetrating fluids: (i) a neutron superfluid 
with current $n_{ n}^{\, \nu}$, and (ii) an electrically charge neutral plasma of electrons and ions that are essentially locked together by the interior magnetic field, 
and whose current $n_{ p}^{\, \nu}$ is carried by protons (although electrons play an important role for electromagnetic effects, their contribution
to the fluid dynamics considered here can be ignored owing to the negligibly small electron mass compared to the proton mass). 
It should be stressed that the neutron superfluid component includes here neutrons that are both bound inside ionic clusters and unbound. The 
two-fluid model can be reformulated in terms of a ``free'' neutron current and a ``confined'' baryon current. However, the specification of 
which neutrons are to be counted as ``free'' or ``confined'' is subject to some degree of arbitrariness. We refer the reader to Ref.~\cite{carter2006d}
for a detailed discussion. We shall ignore here the effects of stratification~\cite{carter2006d}, as well as the small stress anisotropy arising from the 
elasticity of the crust~\cite{carter2006b}, or from strong magnetic fields~\cite{carter2006}. 
However, allowance will be made for the relatively strong entrainment effects between the neutron superfluid and the charged particles. 
As will be shown below, once formulated in the usual space-time decomposition, the fully covariant dynamical equations derived in Ref.~\cite{carter2006d} 
are equivalent to those recently obtained in Refs.~\cite{pethick2010,kobyakov2013} following a more heuristic approach.

\subsection{Equivalence between the convective variational formulation and the traditional approach}
\label{sec:hydro-equiv}

The total force balance equation given by Eq.(2.12) of Ref.~\cite{carter2006d} in the fully covariant approach reads 
\begin{equation}\label{eq:total-force-balance}
 \nabla_\mu T^\mu_{\ \,\nu} = 0\, , 
\end{equation}
where $T^\mu_{\ \,\nu}$ is the material energy-momentum tensor, and we assumed that no external force acts on the system. 
Introducing the total momentum density space vector $g_i=T^0_{\ \,i}$, and decomposing this equation in 
the space-time coordinates leads to Eq.~(10) of Ref.~\cite{kobyakov2013} 
\begin{equation}\label{eq:momentum-conservation}
\frac{\partial}{\partial t} g_i + \nabla_j \Pi^j_{\ \,i} = 0\, , 
 \end{equation}
with $\Pi^j_{\ \,i}\equiv T^j_{\ \,i}$. 
The explicit form of the energy-momentum tensor can be derived using the variational principle and is given by Eq.~(1) of Ref.~\cite{carter2005}: 
\begin{equation}
T^\mu_{\ \,\nu} =\sum_{_{\rm X}}  n_{_{\rm X}}^{\, \mu}\pi^{_{\rm X}}_{\, \nu} +\Psi \delta^\mu_\nu\, , 
\end{equation}
where $\delta^\mu_\nu$ is the Kronecker symbol, and 
\begin{equation}
 \Psi = \Lambda -\sum_{_{\rm X}}  n_{_{\rm X}}^{\, \nu}\pi^{_{\rm X}}_{\, \nu}
\end{equation}
is interpretable as a generalized pressure. 
In the present context, the energy-momentum tensor is given by Eq.(2.11) of Ref.~\cite{carter2006d}. 
In particular, introducing the proton and neutron momenta written as $\mu_i^{ p}$ and $\mu_i^{ n}$ respectively in Ref.~\cite{carter2006d} 
the spatial components of the total momentum density covector and of the energy-momenty tensor read 
\begin{equation}\label{eq:tot-momentum}
 g_i = n_{ n} \mu_i^{ n} + n_{ p} \mu_i^{ p}\, , 
\end{equation}
\begin{equation}
 \Pi^j_{\ \,i} = n_{ p}^j \mu_i^{ p} + n_{ n}^j \mu_i^{ n} +\delta^j_i \Psi\, .
\end{equation}
Under the assumption that the currents are sufficiently small, 
the internal energy density $U$ can quite generally be written as the sum of a purely static part $U_{\rm ins}$ and a dynamical part $U_{\rm dyn}$ (including 
the kinetic contribution) given by Eq.~(2.41) of Ref.~\cite{carter2006d} 
\begin{equation}
U_{\rm dyn} = \frac{1}{2} \left(\mu_i^{ n} n_{ n}^i + \mu_i^{ p} n_{ p}^i \right)\, .
\end{equation}
Likewise, the generalized pressure can be decomposed as 
\begin{equation}
 \Psi = \Psi_{\rm ins} + \Psi_{\rm dyn}\, ,
\end{equation}
where the first term represents a purely static contribution having the form
\begin{equation}\label{eq:ins-pressure}
\Psi_{\rm ins} = n_{ n}\frac{\partial U_{\rm ins}}{\partial n_{ n}} +n_{ p}\frac{\partial U_{\rm ins}}{\partial n_{ p}}  - U_{\rm ins}\, ,
\end{equation}
whereas the second term arises from dynamical effects and is given by 
\begin{equation}\label{eq:dyn-pressure}
\Psi_{\rm dyn} =  - n_{ n}\frac{\partial U_{\rm dyn}}{\partial n_{ n}} -n_{ p}\frac{\partial U_{\rm dyn}}{\partial n_{ p}}  - U_{\rm dyn}\, .
\end{equation}
When taking partial derivatives, it is understood that the relevant variables are the densities $n_{ n}$, $n_{ p}$, and the 
currents $n_{ n}^i$, $n_{ p}^i$. 
 
The superfluidity condition is embedded in Josephson equations, which in the fully covariant approach is given by Eq.~(2.18) of Ref.~\cite{carter2006d}, 
namely\footnote{Since gravity is neglected here, the total momentum covectors $\pi^{_{\rm X}}_{\, \nu}$ reduce to the material momentum covectors 
$\mu^{_{\rm X}}_{\, \nu}$, as can be seen from Eq.~(152) of Ref.~\cite{carter2004} after setting the Newtonian gravitational potential $\phi=0$.}
\begin{equation}\label{eq:Josephson}
\mu^{ n}_{\,\nu} = \hbar \nabla_\nu \varphi^{ n}\, ,
\end{equation}
where $\varphi^{ n}$ is half the phase of the condensate (denoted by $\theta$ in Ref.~\cite{kobyakov2013}), 
and the time component of the 4-momentum covector is interpretable as the opposite of the neutron chemical potential $\mu^{ n}=-\mu^{ n}_{_0}$. 
The latter is expressible as 
\begin{equation}
\mu^{ n} = \frac{\partial U_{\rm ins}}{\partial n_{ n}} -  \frac{\partial U_{\rm dyn}}{\partial n_{ n}}\, ,
\end{equation}
and similarly for the proton chemical potential\footnote{Because of the local electric charge neutrality condition $n_{ p}=n_{ e}$, where $n_{ e}$ 
is the electron number 
density, the electron chemical potential is included in $\mu^p$.}  $\mu^{ p}$. 
As shown in Ref.~\cite{carter2006d}, 
$\mu^{ n}_{\,\nu}$ hence also Eq.~(\ref{eq:Josephson}) are invariant regardless of how the superfluid neutrons are counted. 
Decomposing Eq.~(\ref{eq:Josephson}) into space and time components
yields 
\begin{equation}\label{eq:Josephson3+1}
 \mu_i^{ n} = \hbar \nabla_i \varphi^{ n}\, , \hskip1cm \hbar \frac{\partial \varphi^{ n}}{\partial t} + \mu^{ n} = 0\, . 
\end{equation}
The first condition is traditionally expressed as 
\begin{equation}\label{eq:superfluidity}
V_{{ n}\,i}^{\rm S}  = \frac{\hbar}{m} \nabla_i \varphi^{ n}
\end{equation}
in terms of a ``superfluid velocity'' defined by 
\begin{equation}\label{eq:super-velocity}
V_{{ n}\,i}^{\rm S} \equiv \frac{\mu_i^{ n}}{m}\, , 
\end{equation}
and $m$ is the nucleon mass (we neglect here the small difference between the neutron and proton masses). It can thus be easily seen that 
the superfluidity conditions~(\ref{eq:Josephson3+1}) coincide with Eqs.~(3) and (11) of Ref.~\cite{kobyakov2013}. 

As stressed by Carter (see, e.g. Ref.~\cite{carter1994}), $V_i^{\rm S}$ does not represent the true physical velocity 
of the neutron superfluid, denoted here by $v_{ n}^{\,i}$. The different nature of $V_{n\,i}^{\rm S}$ and $v_{ n}^{\,i}$ 
appears very clearly in the 4-dimensionally covariant approach, see Eq.~(\ref{eq.full.momentum}). It should be stressed that 
in Newtonian spacetime, vectors (such as the true velocity) and covectors (such as the ``superfluid velocity'') are intrinsically 
different objects due to the absence of a metric tensor (indices cannot be raised or lowered). 
Although the mathematical distinction between velocity and momentum seems to disappear in classical 
hydrodynamics formulated in the usual space time decomposition with Cartesian coordinates, this is no longer the case when 
dealing with superfluid systems. In particular, in superfluid mixtures such as helium-3 and helium-4, the different superfluids 
are generally mutually coupled by (nondissipative) entrainment effects whereby the true velocity of one species is not aligned with 
the corresponding ``superfluid velocity'' even in Cartesian coordinates~\cite{andreev1976}. In the present context, the neutron  and proton 
momenta can thus be generally written as
\begin{equation}\label{eq:momentum-velocity}
 \mu_i^{ n} =\gamma_{ij} ({\cal K}^{ nn}\,n_{ n}^{\,j} +{\cal K}^{ np}\, n_{ p}^{\,j})\, , \hskip1cm  
 \mu_i^{ p} =\gamma_{ij} ({\cal K}^{ np}\,n_{ n}^{\,j} +{\cal K}^{ pp}\, n_{ p}^{\,j})
\end{equation}
where $\gamma_{ij}$ denotes the space metric. The coefficients ${\cal K}^{ nn}$, ${\cal K}^{ np}$, and ${\cal K}^{ pp}$ 
are not independent, but must satisfy the following conditions imposed by Galilean invariance~\cite{carter2006d}: 
\begin{equation}\label{eq:Galilean}
{\cal K}^{ nn}\,n_{ n} +{\cal K}^{ np}\, n_{ p}=m\, , \hskip1cm  
{\cal K}^{ np}\,n_{ n} +{\cal K}^{ pp}\, n_{ p}=m\, .
\end{equation}
Entrainment effects can thus be described by only one coefficient, for instance ${\cal K}^{ np}$. 
Indeed, the dynamical contribution $U_{\rm dyn}$ to the internal energy density can be decomposed into a kinetic term 
\begin{equation}\label{eq:kin-energy}
U_{\rm kin} = \frac{1}{2} m \left( n_{ n} v_{ n}^2 + n_{ p} v_{ p}^2 \right)\, ,
\end{equation}
and an entrainment term 
\begin{equation}\label{eq:ent-energy}
U_{\rm ent} = \frac{1}{2} \bar\rho_{ np}  \bar v_{ np}^2 \, ,
\end{equation}
where 
$\bar v_{ np}^i = v_{ n}^i - v_{ p}^i$ 
is the relative velocity, and $\bar\rho_{ np} = -n_{ n}n_{ p}{\cal K}^{ np}$~\cite{carter2006d}. 
Imposing $U_{\rm dyn}>0$, as required for the stability of the static configuration, entails the additional 
constraint
\begin{equation}\label{eq:stability}
\bar\rho_{ np}> -  x_{ p} (1-x_{ p})\rho\, , 
\end{equation}
where $\rho = m(n_{ n}+n_{ p})$ is the  mass density, and $x_{ p}=n_{ p}/(n_{ n}+n_{ p})$ is the proton fraction. 
Inserting $U_{\rm dyn}=U_{\rm kin}+U_{\rm ent}$ 
in Eq.~(\ref{eq:dyn-pressure}) using Eqs.~(\ref{eq:kin-energy}) and (\ref{eq:ent-energy}), the generalized pressure is expressible as 
\begin{equation}
\Psi_{\rm dyn} = \frac{1}{2} \bar v_{ np}^2\left(\bar\rho_{ np} - n_{ n}\frac{\partial \bar\rho_{ np}}{\partial n_{ n}} 
-n_{ p}\frac{\partial \bar\rho_{ np}}{\partial n_{ p}}\right)\, .
\end{equation}

In Ref.~\cite{kobyakov2013} the neutron current (denoted by $j_{ n}^i$) was alternatively expressed in terms of the ``superfluid velocity'' and the 
proton velocity $v_{ p}^i$ in a more traditional form as 
\begin{equation}\label{eq:velocity-momentum}
n_{ n}^i = n_{ n}^{\rm S} V_{ n}^{{\rm S}\,i} +  n_{ n}^{\rm N} v_{ p}^i\, , 
\end{equation}
where $n_{ n}^{\rm S}$ and $n_{ n}^{\rm N}$ were referred to as the ``superfluid'' and ``normal'' neutron density 
respectively. Comparing Eqs.~(\ref{eq:momentum-velocity}) and (\ref{eq:velocity-momentum}), we find
\begin{equation}
 n_{ n}^{\rm S} = \frac{m}{{\cal K}^{ nn}}\, ,
\end{equation}
or equivalently
\begin{equation}
 n_{ n}^{\rm S} = n_{ n}\left(1+\frac{\bar\rho_{ np}}{\rho_{ n}}\right)^{-1}\, .
\end{equation}
The stability condition~(\ref{eq:stability}) can thus be written as 
\begin{equation}
\frac{n_{ n}^{\rm S}}{n_{ n}} < \frac{1}{1-x_{ p}}\, .
\end{equation}
It is easily seen that the relations~(\ref{eq:Galilean}) can be equivalently expressed as
\begin{equation}
 n_{ n}^{\rm N}+n_{ n}^{\rm S}=n_{ n}\, .
\end{equation}
With these notations, the total momentum density coincides with Eq.~(3.8) of Ref.~\cite{pethick2010}, namely 
\begin{equation}\label{eq:tot-momentumS}
 g_i = m n_{ n}^{\rm S} V_{{ n}\,i}^{\rm S} + (\rho-m n_{ n}^{\rm S})  v_{{ p}\,i}\, .
\end{equation}
The dynamical energy becomes 
\begin{equation}
 U_{\rm dyn} = \frac{1}{2} m n_{ n}^{\rm S}V_{{ n}}^{\rm S\, 2} + \frac{1}{2} (\rho-m n_{ n}^{\rm S}) v_{ p}^2 \, .
\end{equation}

Entrainment can be equivalently formulated in terms of dynamical effective masses~\cite{carter2006d}. In the crust rest frame 
($v_{ p}^i=0$), the neutron momentum can thus be written as $\mu_i^{ n}=\gamma_{ij}m_{ n}^\star v_{ n}^j$. 
Alternatively, a second kind of effective mass can be defined by expressing $\mu_i^{ n}=\gamma_{ij}m_{ n}^\sharp v_{ n}^j$ 
in the crust \emph{momentum} rest frame ($\mu_i^{ p}=0$). 
Using Eqs.~(\ref{eq:momentum-velocity}), we obtain 
\begin{equation}
m_{ n}^\star=n_{ n}{\cal K}^{ nn}\, , \hskip1cm m_{ n}^\sharp=n_{ n}\frac{{\cal K}^{ nn}{\cal K}^{\rm pp}-({\cal K}^{ np})^2}{{\cal K}^{ pp}} \, .
\end{equation} 
Although this formulation could provide a more intuitive interpretation of entrainment, it is not devoid of ambiguity. Indeed, 
these dynamical effective masses are found to depend on how ``free'' and ``confined nucleons are defined. More importantly, different 
definitions of effective masses have been introduced in various contexts to characterize different physical aspects. In particular, the 
dynamical effectives masses above should not be confused with those introduced in microscopic many-body theories (see, e.g. Ref.~\cite{chamel2006}). 

To be complete, the system of dynamical equations Eqs.~(\ref{eq:total-force-balance}) and (\ref{eq:Josephson}) 
for the currents $n_{ n}^{\, \nu}$, $n_{ p}^{\, \nu}$, and the phase $\varphi^{ n}$ must be supplemented with a further condition. 
On sufficiently short dynamical timescales, neutrons and protons can be reasonably assumed to be separately conserved, which can be 
covariantly expressed as
\begin{equation}\label{eq:continuity}
 \nabla_{\!\nu} n_{ n}^{\,\nu}=0\, , \hskip1cm \nabla_{\!\nu}n_{ p}^{\,\nu}=0\, .
\end{equation}
In the usual space-time decomposition, these equations become
\begin{equation}\label{eq:continuity3+1}
 \frac{\partial n_{ n}}{\partial t} + \nabla_i n_{ n}^i = 0\, , \hskip1cm  \frac{\partial n_{ p}}{\partial t} + \nabla_i n_{ p}^i = 0\, , 
\end{equation}
which are equivalent to Eqs.(8) and (9) of Ref.~\cite{kobyakov2013}. In fact, only one of Eqs.~(\ref{eq:continuity}) needs to be considered 
if the conservation of the full energy-momentum tensor is imposed, see Eqs.~(\ref{eq:total-force-balance}). Alternatively, the dynamical evolution 
can be fully determined by Eqs.~(\ref{eq:momentum-conservation}), (\ref{eq:Josephson3+1}), and (\ref{eq:continuity3+1}). Finally, let us remark 
that the hydrodynamical equations could have been equivalently derived from Eq.~(\ref{eq.hydro.covariant}) with $f^{_{\rm X}}_{\,\nu}=0$. 

\subsection{Low-energy collective excitations of superfluid neutron-star crusts}
\label{sec:hydro-excitations}

In Refs.~\cite{pethick2010,kobyakov2013}, the dynamical equations~(\ref{eq:momentum-conservation}), (\ref{eq:Josephson3+1}), and (\ref{eq:continuity3+1})
were further simplified considering small perturbations 
against an initially static background. The particle number conservation Eqs.~(\ref{eq:continuity3+1}) thus become to first order 
\begin{equation}\label{eq:linearized-continuity}
 \frac{\partial \delta n_{ n}}{\partial t} +  n_{ n}^{\rm S}\nabla_i \delta V_{ n}^{{\rm S}\, i} + n_{ n}^{\rm N}\nabla_i\delta v_{ p}^i = 0\, , \hskip1cm  \frac{\partial \delta n_{ p}}{\partial t} + n_{ p}\nabla_i \delta v_{ p}^i = 0\, , 
\end{equation}
using Eq.~(\ref{eq:velocity-momentum}). To linearize the momentum conservation Eq.~(\ref{eq:momentum-conservation}), we make use of the generalized 
Gibbs-Duhem identity, see Eq.(151) of Ref.~\cite{carter2004} (ignoring here gravity), namely
\begin{equation}
 \delta \Psi = - \sum_{_{\rm X}} n_{_{\rm X}}^{\, \nu}\delta \mu^{\rm X}_{\,\nu}\, , 
\end{equation}
which in the present context reduces to Eq.~(19) of Ref.~\cite{kobyakov2013}
\begin{equation}\label{eq:deltaPsi}
 \delta \Psi = n_{ n} \delta \mu^{ n} + n_{ p} \delta \mu^{ p}\, . 
\end{equation}
Using Eqs.~(\ref{eq:tot-momentum}), (\ref{eq:super-velocity}), (\ref{eq:tot-momentumS}), and (\ref{eq:deltaPsi}), the momentum conservation is found 
to coincide with Eq.~(21) of Ref.~\cite{kobyakov2013}
\begin{equation}\label{eq:linearized-momentum}
 m n_{ n}^{\rm S} \frac{\partial \delta V_{{ n}\, i}^{\rm S}}{\partial t}+m(n_{ n}^{\rm N}+n_{ p})\frac{\partial \delta v_{{ p}\,i}}{\partial t}
 +  n_{ n} \nabla_i \delta \mu^{ n} + n_{ p} \nabla_i \delta \mu^{ p} = 0\, . 
\end{equation}
The linearized version of the Josephson equations~(\ref{eq:Josephson3+1}) leads to the same equation as Eq.~(23) of Ref.~\cite{kobyakov2013}:  
\begin{equation}\label{eq:linearized-Josephson3+1}
m \frac{\partial \delta V_{{ n}\,i}^{\rm S}}{\partial t} + \nabla_i \delta \mu^{ n} = 0\, . 
\end{equation}
Rearranging Eqs.~(\ref{eq:linearized-momentum}) and (\ref{eq:linearized-Josephson3+1}) as in Ref.~\cite{kobyakov2013} yields
\begin{equation}\label{eq:linearized-momentum-bis}
m(n_{ n}^{\rm N}+n_{\rm p})\frac{\partial \delta v_{{ p}\,i}}{\partial t}
 +  n_{ n}^{\rm N} \nabla_i \delta \mu^{ n} + n_{ p} \nabla_i \delta \mu^{ p} = 0\, . 
\end{equation}

Let us consider perturbations in the form of plane waves that vary in space and time as $\exp[{\rm i}(q_i x^i -\omega t)]$, where $q_i$ are the coordinates of the 
wave vector and $\omega$ is the angular frequency. In the long wavelength limit $q\rightarrow0$, such perturbations have a soundlike dispersion 
relation of the form $\omega=v q$, where $v$ is the corresponding propagation speed. The properties of these modes are of particular importance for 
studying the thermal evolution of neutron-star crusts~\cite{chamel2013c}. 
In the minimal model that we consider here, the modes are purely longitudinal. 
In the absence of the neutron superfluid (as in the outer crust of a neutron star), these modes are lattice vibrations propagating at the speed 
\begin{equation}
\label{eq:lattice-vibration-speed}
v^0_\ell = \sqrt{\frac{\widetilde K}{\rho_{\rm I}}}\, , 
\end{equation}
where $\widetilde K$ is the bulk modulus of the electron-ion system, given by\footnote{Let us recall that the chemical potential $\mu^{ p}$ includes 
the contribution of electrons.}
\begin{equation}
 \widetilde K=n_{ p}^2 \frac{\partial \mu^{ p}}{\partial n_{ p}}\, , 
\end{equation}
and $\rho_{\rm I}$ is the ion mass density. A pure neutron superfluid with density $n^{\rm f}_{ n}$ could be subject to Bogoliubov-Anderson perturbations 
with a speed 
\begin{equation}\label{eq:Bogoliubov-Anderson-speed}
v^0_\phi=\sqrt{\frac{n^{\rm f}_{ n}}{m}\frac{\partial\mu^{ n}}{\partial n^{\rm f}_{ n}}}\, . 
\end{equation}
In the shallowest region of the inner crust, the neutron superfluid is very dilute and weakly coupled to the electron-ion plasma. The excitations modes 
of the combined system can still be decomposed into lattice vibrations and Bogoliubov-Anderson sound mode with $n_{ n}^{\rm f}$ the density of ``free'' 
neutrons. The speeds of these two modes are such that 
$v^0_\phi\ll v^0_\ell$. With increasing depth, the speeds of these modes are changed due to entrainment effects. In particular, lattice vibrations 
are accompanied by motions of the neutron superfluid so that their speed is reduced~\cite{chamel2013c}: 
\begin{equation}
\label{eq:lattice-vibration-speed-entrainment}
v_\ell = v_\ell^0 \sqrt{\frac{\rho_{\rm I}}{m(n_{ p}+n_{ n}^{\rm N})}}\leq v_\ell^0\, .
\end{equation}
Likewise, because the electron-ion plasma is entrained by the neutron superfluid, the speed of the Bogoliubov-Anderson mode is increased~\cite{chamel2013c}:     
\begin{equation}
v_\phi=v_\phi^0 \sqrt{\frac{n_{ n}^{\rm S}}{n^{\rm f}_{ n}}}\geq v_\phi^0\, . 
\end{equation}
With further compression, the speeds of the two modes become comparable and mix. The speeds can be determined from 
Eqs.~(\ref{eq:linearized-continuity}), (\ref{eq:linearized-Josephson3+1}), (\ref{eq:linearized-momentum-bis}) leading to 
an equation of the form~\cite{chamel2016}
\begin{equation}\label{eq:dispersion-relation}
(v^2-v_\phi^2)(v^2-v_\ell^2)=g^2_{\rm mix}v^2 +g^4\, .
\end{equation}
The mixing between the modes is characterized by the parameters $g_{\rm mix}$ and $g$, given by 
\begin{equation}
g_{\rm mix}=\sqrt{\frac{n_n^{\rm N} (2 L+E_{ nn} n_n^{\rm N})}{m(n_{ p}+n_{ n}^{\rm N})}}\, , \hskip 0.5cm g=\left(\frac{L^2 n_n^{\rm S}}{m^2(n_{ p}+n_{ n}^{\rm N})}\right)^{1/4}\, ,
\end{equation}
where 
\begin{equation}
 L=n_{ p}\frac{\partial \mu^{ n}}{\partial n_{ p}}\, , \hskip0.5cm E_{ nn}=\frac{\partial\mu^{ n}}{\partial n_{ n}}\, .
\end{equation}
Note that Eq.~(\ref{eq:dispersion-relation}) is identical to Eq.~(34) from Ref.~\cite{kobyakov2013} although it is expressed here in a slightly different form. 
The two solutions of Eq.~(\ref{eq:dispersion-relation}) are given by~\cite{chamel2016} 
\begin{equation}
 v_{\pm} = \frac{V}{\sqrt{2}}\sqrt{1\pm\sqrt{1-\frac{4 v_\ell^2 v_\phi^2}{V^4}+\frac{4g^4}{V^4}}}\, ,
\end{equation}
where $V = \sqrt{v_\ell^2 + v_\phi^2 +g_{\rm mix}^2}$. In the deep region of the inner crust, most nucleons consist of superfluid neutrons so that the two modes 
ressemble lattice vibrations and Bogoliubov-Anderson excitations, with speeds $v_- \sim v_\ell \ll v_+ \sim v_\phi$.  

In the non-superfluid phase, any relative motion between the neutron liquid and the crust will be damped by viscosity to the effect that 
ions, electrons and neutrons will be essentially comoving. In this case, the Josephson's equation~(\ref{eq:linearized-Josephson3+1}) have to 
be replaced by the condition $\delta v_{ n}^i = \delta v_{ p}^i$. Only one longitudinal mode corresponding to ordinary 
hydrodynamic sound will persist and its speed will be given by
\begin{equation}
c_s =\sqrt{\frac{K}{\rho}}\, ,
\end{equation}
where $\rho$ is the total mass density of the crust and $K$ is the total bulk modulus
\begin{equation}
 K=\widetilde K + 2 n_{ n} L + n_{ n}^2 E_{ nn}\, .
\end{equation}

This analysis illustrates the importance of entrainment effects on the dynamics of neutron-star crusts. The complete characterization of the 
modes (and more generally the complete determination of the superfluid dynamics) requires the specification of the static internal energy 
density $U_{\rm ins}$, as well as of the superfluid density $n_{ n}^{\rm S}$. These microscopic ingredients can be calculated using the 
nuclear EDF theory.

\section{Microscopic description of neutron-star crusts and origin of entrainment}
\label{sec:micro}
\subsection{Nuclear energy density functional theory}
\label{sec:micro-EDF}

The density functional theory has been very successfully employed in a wide variety of fields, from chemistry to condensed matter physics. 
A somehow similar approach called the nuclear EDF theory, has been developed in nuclear physics (see, e.g. 
Ref.~\cite{duguet14} for a recent review of this formalism). 

In this theory, the energy $E$ of a many-nucleon system is expressed as a universal functional of the so called normal and abnormal density 
matrices~\cite{doba84,doba96} defined by
\begin{equation}
n_q(\pmb{r}, \sigma; \pmb{r^\prime}, \sigma^\prime) = <\Psi|c_q(\pmb{r^\prime},\sigma^\prime)^\dagger c_q(\pmb{r},\sigma)|\Psi>\, ,
\end{equation}
\begin{equation}
\widetilde{n_q}(\pmb{r}, \sigma; \pmb{r^\prime}, \sigma^\prime) = -\sigma^\prime <\Psi|c_q(\pmb{r^\prime},-\sigma^\prime) c_q(\pmb{r},\sigma)|\Psi>\, , 
\end{equation}
respectively where $|\Psi>$ is the many-body wave function, $c_q(\pmb{r},\sigma)^\dagger$ and $c_q(\pmb{r},\sigma)$ are the creation and destruction operators 
for nucleons of charge type $q$ ($q=n,p$ for neutron, proton respectively)  at position $\pmb{r}$ with spin $\sigma$. 
The abnormal density matrix characterizes the pairing of nucleons (see, e.g., the discussion in Ref.~\cite{doba96}). 
The normal and abnormal density matrices are usually expressed in terms of independent quasi-particle (q.p.) states characterized by two-component 
wavefunctions $\psi^{(q)}_{1i}(\pmb{r}, \sigma)$ and $\psi^{(q)}_{2i}(\pmb{r}, \sigma)$, as  
\begin{equation}
 n_q(\pmb{r}, \sigma; \pmb{r^\prime}, \sigma^\prime) =
\sum_{i(q)}\psi^{(q)}_{2i}(\pmb{r}, \sigma)\psi^{(q)}_{2i}(\pmb{r^\prime}, \sigma^\prime)^* \, ,
\end{equation}
and
\begin{equation}
\widetilde{n_q}(\pmb{r}, \sigma; \pmb{r^\prime}, \sigma^\prime) =
-\sum_{i(q)}\psi^{(q)}_{2i}(\pmb{r}, \sigma)
\psi^{(q)}_{1i}(\pmb{r^\prime}, \sigma^\prime)^*=-
\sum_{i(q)}\psi^{(q)}_{1i}(\pmb{r}, \sigma)\psi^{(q)}_{2i}(\pmb{r^\prime},
\sigma^\prime)^* \, ,
\end{equation}
where the index $i$ represents the set of suitable quantum numbers and the symbol $*$ denotes complex conjugation. 
The ground-state energy of the system is determined by minimizing the energy $E$ with respect to $\psi^{(q)}_{1i}(\pmb{r}, \sigma)$ and 
$\psi^{(q)}_{2i}(\pmb{r}, \sigma)$ under the constraint of fixed numbers of neutrons and protons.

The main limitation of the EDF theory stems from the energy functional itself, whose exact form is unknown. For this reason, various 
phenomenological functionals have been proposed. They have been traditionally obtained from density-dependent effective nucleon-nucleon 
interactions in the framework of the self-consistent ``mean-field'' methods~\cite{bhr03}. Although such a formulation imposes stringent 
restrictions on the form of the EDF, it guarantees the cancellation of the internal energy in the limiting case of one nucleon~\cite{cha10}. 
On the other hand, the EDFs may still be contaminated by many-body self-interactions errors (see, e.g. Ref.~\cite{duguet14}). 
The EDFs reduce to a semi-local form for zero-range effective interactions. Such interactions have been widely employed since they allow 
for very fast numerical computations. In particular, the most popular effective interactions are of the Skyrme type~\cite{bhr03}
\begin{eqnarray}\label{eq:skyrme}
v(\pmb{r}_{1},\pmb{r}_{2}) & = & t_0(1+x_0 P_\sigma)\delta({\pmb{r}_{12}})+\frac{1}{2} t_1(1+x_1 P_\sigma)\frac{1}{\hbar^2}\left[p_{12}^2\,
\delta({\pmb{r}_{12}}) +\delta({\pmb{r}_{12}})\, p_{12}^2 \right]\nonumber\\
&+&t_2(1+x_2 P_\sigma)\frac{1}{\hbar^2}\pmb{p}_{12}\cdot\delta(\pmb{r}_{12})\,\pmb{p}_{12}+\frac{1}{6}t_3(1+x_3 P_\sigma)n(\pmb{r})^\alpha\,\delta(\pmb{r}_{12})
\nonumber\\
& +&\frac{\rm i}{\hbar^2}W_0(\pmb{\hat\sigma_1}+\pmb{\hat\sigma_2})\cdot
\pmb{p}_{12}\times\delta(\pmb{r}_{12})\,\pmb{p}_{12} 
\, , 
\end{eqnarray}
where $\pmb{r}_{12} = \pmb{r}_1 - \pmb{r}_2$, $\pmb{r} = (\pmb{r}_1 + 
\pmb{r}_2)/2$, $\pmb{p}_{12} = - {\rm i}\hbar(\pmb{\nabla}_1-\pmb{\nabla}_2)/2$
is the relative momentum, $\pmb{\hat\sigma_1}$ and $\pmb{\hat\sigma_2}$ are Pauli spin matrices, 
 $P_\sigma$ is the two-body spin-exchange operator, and $n(\pmb{r})$ denotes the average nucleon number 
density. Nuclear pairing is generally treated using a different effective interaction of the form (see, e.g. Ref.~\cite{cha10a} and 
references therein) 
\begin{equation}\label{eq:pairing}
 v(\pmb{r}_{1},\pmb{r}_{2})=\frac{1}{2}(1- P_\sigma) v^{\pi\,q}[n_n(\pmb{r}),n_p(\pmb{r})]\delta(\pmb{r}_{12}) \, ,
\end{equation}
 where $n_n(\pmb{r})$ and $n_p(\pmb{r})$ denote the average neutron and proton number densities respectively. Only pairing between nucleons of the same charge 
 state is considered here. Because of the zero range, the pairing interaction must be regularized. This is usually achieved by introducing an energy cutoff 
 (for a review of the various prescriptions, see for instance Ref.~\cite{dug05}). 
 
 With these kinds of zero-range interactions, the energy $E$ can be expressed as 
  \begin{equation}
  E=E_{\rm kin}+E_{\rm Coul}+E_{\rm Sky}+E_{\rm pair} \, ,
 \end{equation}
 where $E_{\rm kin}$ is the kinetic energy, $E_{\rm Coul}$ is the Coulomb energy, $E_{\rm Sky}$ is the Skyrme nuclear energy, and $E_{\rm pair}$ is the nuclear 
 pairing energy. 
Assuming the system to be invariant under time reversal, the ground-state energy depends on the local normal and abnormal nucleon number densities
\begin{equation}\label{eq:rhoq}
n_q(\pmb{r}) = \sum_{\sigma=\pm 1}n_q(\pmb{r}, \sigma; \pmb{r}, \sigma)\, ,
\end{equation}
\begin{equation}\label{eq:rhoq}
\widetilde{n_q}(\pmb{r}) = \sum_{\sigma=\pm 1}\widetilde{n_q}(\pmb{r}, \sigma; \pmb{r}, \sigma)\, ,
\end{equation}
the kinetic densities
\begin{equation}\label{eq:tauq}
\tau_q(\pmb{r}) = \sum_{\sigma=\pm 1}\int\,{\rm d}^3\pmb{r^\prime}\,\delta(\pmb{r}-\pmb{r^\prime}) \pmb{\nabla}\cdot\pmb{\nabla^\prime}
n_q(\pmb{r}, \sigma; \pmb{r^\prime}, \sigma)\, , 
\end{equation}
and the spin-current vector densities
\begin{eqnarray}\label{eq:Jq}
\pmb{J}_q(\pmb{r}) &=& -{\rm i}\sum_{\sigma,\sigma^\prime=\pm1}\int\,{\rm d}^3\pmb{r^\prime}\,\delta(\pmb{r}-\pmb{r^\prime}) 
\pmb{\nabla} n_q(\pmb{r}, \sigma; \pmb{r^\prime},
\sigma^\prime) \times  \pmb{\sigma}_{\sigma^\prime\sigma}   \nonumber \\
&=&{\rm i}\sum_{\sigma,\sigma^\prime=\pm1}\int\,{\rm d}^3\pmb{r^\prime}\,\delta(\pmb{r}-\pmb{r^\prime}) 
\pmb{\nabla^\prime} n_q(\pmb{r}, \sigma; \pmb{r^\prime},
\sigma^\prime) \times  \pmb{\sigma}_{\sigma^\prime\sigma}\, , 
\end{eqnarray}
where $ \pmb{\sigma}_{\sigma\sigma^\prime}$ denotes the components of the Pauli spin matrices. 
The energy minimization leads to the Hartree-Fock-Bogoliubov (HFB) equations\footnote{These equations are 
also called Bogoliubov-de Gennes equations in condensed matter physics.}
\begin{eqnarray}
\label{eq:HFB}
\sum_{\sigma^\prime=\pm1}
\begin{pmatrix} h_q(\pmb{r} )_{\sigma \sigma^\prime} & \Delta_q(\pmb{r}) \delta_{\sigma \sigma^\prime} \\ \Delta_q(\pmb{r}) \delta_{\sigma \sigma^\prime} & -h_q(\pmb{r})_{\sigma \sigma^\prime} 
 \end{pmatrix}\begin{pmatrix} 
\psi^{(q)}_{1i}(\pmb{r},\sigma^\prime) \\ \psi^{(q)}_{2i}(\pmb{r},\sigma^\prime) \end{pmatrix} = \nonumber\\
\begin{pmatrix} E_i+\mu^q & 0 \\ 0 & E_i-\mu^q \end{pmatrix}
\begin{pmatrix} \psi^{(q)}_{1i}(\pmb{r},\sigma) \\ \psi^{(q)}_{2i}(\pmb{r},\sigma) \end{pmatrix}\, ,
\end{eqnarray}
where $E_i$ denotes the energy of the q.p. state $i$, and the chemical potentials $\mu^q$ introduced as Lagrange multipliers to impose the constraints on 
the fixed numbers $N_q$ of nucleons are determined from the condition 
\begin{equation}\label{eq:Nq-HFB}
N_q = \sum_i \sum_\sigma \int d^3r\, \vert \psi^{(q)}_{2i}(\pmb{r},\sigma)\vert^2 \, . 
\end{equation}
The single-particle (s.p.) Hamiltonian $h_q(\pmb{r} )_{\sigma \sigma^\prime}$ is given by 
\begin{equation}
h_q(\pmb{r})_{\sigma\sigma^\prime} \equiv -\pmb{\nabla}\cdot
B_q(\pmb{r})\pmb{\nabla}\, \delta_{\sigma\sigma^\prime}
+ U_q(\pmb{r}) \delta_{\sigma\sigma^\prime}
-{\rm i}\pmb{W_q}(\pmb{r}) \cdot\pmb{\nabla}\times\pmb{\hat\sigma}_{\sigma\sigma^\prime}\, , 
\end{equation}
with the s.p. fields defined by the functional derivatives of the energy
\begin{equation}
 B_q(\pmb{r}) =
\frac{\delta E}{\delta\tau_q(\pmb{r})}\, ,
\hskip0.5cm
U_q(\pmb{r})=\frac{\delta E}{\delta n_q(\pmb{r})}\, ,
\hskip0.5cm
\pmb{W}_q(\pmb{r})=\frac{\delta E}{\delta \pmb{J}_q(\pmb{r})}  \, .
\end{equation}
The pairing potential is defined by 
\begin{equation}
\Delta_q(\pmb{r})=\frac{\delta E}{\delta \widetilde{n_q}(\pmb{r})}=\frac{1}{2}v^{\pi q} [n_n(\pmb{r}),n_p(\pmb{r})]\widetilde{n_q}(\pmb{r}) \, .
\end{equation}
Expressions for these fields can be found for instance in Ref.~\cite{cha08}. 

In the absence of pairing, the HFB equations reduce to the Hartree-Fock (HF) equations
\begin{equation}\label{eq:HF}
\sum_{\sigma^\prime=\pm1}\,  h_q(\pmb{r})_{\sigma\sigma^\prime} \varphi^{(q)}_{i}(\pmb{r},\sigma^\prime) = \varepsilon^{(q)}_i \varphi^{(q)}_{i}(\pmb{r},\sigma)\, , 
\end{equation}
and $\varepsilon^{(q)}_i$ is the energy of the s.p. state $i$ characterized by the s.p. wavefunction $\varphi^{(q)}_{i}(\pmb{r},\sigma)$. 
The so called BCS approximation consists in expressing the HFB equations in the HF basis\footnote{The pairing contributions 
to $h_q$ are typically very small, and therefore often neglected.}, and neglecting the off-diagonal matrix elements of the pairing potential. 
Adopting the usual phase convention, the solutions of the HFB Eqs.~(\ref{eq:HFB}) are thus given by 
\begin{equation}\label{eq:psi-bcs}
 \psi^{(q)}_{1i}(\pmb{r},\sigma) = U^{(q)}_i \varphi^{(q)}_i(\pmb{r},\sigma)\, , \hskip0.5cm \psi^{(q)}_{2i}(\pmb{r},\sigma) = V^{(q)}_i\,\varphi^{(q)}_{i}(\pmb{r},\sigma)\, ,
 \end{equation}
 \begin{equation}\label{eq:Uk-Vk}
U^{(q)}_i = \frac{1}{\sqrt{2}}\Biggl[1+\frac{\varepsilon^{(q)}_i-\mu^q}{E^{(q)}_i}\Biggr]^{1/2}\, , \hskip0.5cm  V^{(q)}_i = -\frac{1}{\sqrt{2}}\Biggl[1-\frac{\varepsilon^{(q)}_i-\mu^q}{E^{(q)}_i}\Biggr]^{1/2}\, , 
 \end{equation} 
\begin{equation}\label{eq:Eqp}
 E^{(q)}_i=\sqrt{(\varepsilon^{(q)}_i-\mu^q)^2+\Delta^{(q)2}_i}\, .
\end{equation}
The condition~(\ref{eq:Nq-HFB}) reduces to 
\begin{equation}\label{eq:Nq-BCS}
N_q = \sum_i V^{(q) 2}_i \, .
\end{equation}
The pairing gaps $\Delta^{(q)}_i$ are determined by the BCS equations 
\begin{equation}\label{eq:BCS-gap}
 \Delta^{(q)}_{i} = - \frac{1}{2} \sum_{j}V^{(q)}_{ij}  
 \frac{\Delta^{(q)}_{j}}{E^{(q)}_{j} } \, ,
\end{equation}
\begin{equation}
V^{(q)}_{i j} = \frac{1}{2} \sum_{\sigma} \int d^3 r 
\vert \varphi^{(q)}_{i}(\pmb{r},\sigma)\vert^2 v^{\pi q} [n_n(\pmb{r}),n_p(\pmb{r})]\vert\varphi^{(q)}_{j}(\pmb{r},\sigma)\vert^2  \, .
\end{equation}
The BCS ansatz actually provides an exact solution of the HFB equations for homogeneous systems. 

Depending on the choice of boundary conditions, the HFB or HF(+BCS) equations can describe atomic nuclei, neutron-star crusts, or 
homogeneous nuclear matter as in the core of neutron stars. 

\subsection{Application to neutron-star crusts}
\label{sec:micro-crust}

Assuming that the crust of a neutron star consists of a perfect crystal, the neutron and proton q.p. states are characterized by a band index $\alpha$ and 
a Bloch wave vector $\pmb{k}$. The corresponding q.p. wavefunctions must obey the following boundary conditions~\cite{mat76}
\begin{eqnarray}\label{eq:Bloch}
 \psi^{(q)}_{1\alpha\pmb{k}}(\pmb{r}+\pmb{\ell}, \sigma)=\exp({\rm i} \pmb{k}\cdot\pmb{\ell})\,\psi^{(q)}_{1\alpha\pmb{k}}(\pmb{r}, \sigma)\nonumber\\
 \psi^{(q)}_{2\alpha\pmb{k}}(\pmb{r}+\pmb{\ell}, \sigma)=\exp({\rm i} \pmb{k}\cdot\pmb{\ell})\,\psi^{(q)}_{2\alpha\pmb{k}}(\pmb{r}, \sigma) 
\end{eqnarray}
for any lattice translation vector $\pmb{\ell}$, as imposed by the Floquet-Bloch theorem (see, e.g., Ref.~\cite{ash76}). Solving the HFB equations~(\ref{eq:HFB}) fully 
self-consistently with Bloch boundary conditions~(\ref{eq:Bloch}) represents a computationally extremely onerous task, even in the case of semilocal EDFs. 
The main reason stems from the fact that the structure and the composition of the crust of a neutron star are not a priori known, contrary to the case 
of electrons in ordinary materials, or cold atoms in optical lattices. It is generally assumed that during the formation of a neutron star in gravitational 
core-collapse supernova explosions and the subsequent cooling, the dense stellar matter undergoes all kinds of electroweak and nuclear reactions until it 
eventually becomes cold and fully ``catalyzed''~\cite{hw58,htww65}. Determining the ground state of any layer of the crust of a neutron star at some given pressure 
thus requires to solve the coupled HFB equations~(\ref{eq:HFB}) for both neutrons and protons (together with Poisson's equation for the Coulomb electrostatic potential) 
considering all possible compositions and crystal lattice structures\footnote{In principle, one should also solve the density-functional theory equations 
for electrons. But in the extreme environment of neutron stars it is usually a very good approximation to treat electrons as an ideal relativistic Fermi gas.}. 
Such calculations must be repeated for all pressures prevailing in the crust, from $P=0$ at the surface to $\sim 4-7\times 10^{32}$ dyn~cm$^{-2}$ at the crust-core 
boundary. 

In the outermost region of the crust at pressures $P\lesssim 8\times 10^{29}$ dyn~cm$^{-2}$, the determination of the equilibrium structure 
is considerably simplified since all nucleons are bound inside nuclei that are very far apart from each other. In this case, the q.p. states are essentially 
independent of $\pmb{k}$ and the HFB equations can thus be solved for a single isolated nucleus (whose mass is given by the HFB energy $E$ divided by the square 
of the speed of light, i.e. $E/c^2$). Since any given layer of the outer crust is usually made of only one type of nuclei due to gravitational 
settling, the crystal structure is expected to be body-centered cubic (see, e.g. Ref.~\cite{chamel2016b} and references therein). The composition predicted by recent nuclear mass models can be found in Refs.~\cite{wolf2013,pearson2011,kreim2013,bcpm2015,utama2016,chamel2017}. 

The determination of the 
equilibrium structure of the inner regions of neutron-star crusts is much more challenging due to the presence of unbound neutrons. For this reason, following 
the pioneer work of Negele and Vautherin in 1973~\cite{nv73}, most HFB calculations (see, e.g. Ref.~\cite{margueron2012}) have been performed using an 
approximation introduced by Eugene Wigner and Frederick Seitz in 1933 in the context of electrons in metals~\cite{wigner1933}. Namely, the Wigner-Seitz 
or Voronoi cell of the lattice (defined by the set of points that are closer to a given lattice site than to any other) is approximated by a sphere of equal 
volume, and the Bloch boundary conditions (\ref{eq:Bloch}) are replaced by the requirement that the neutron and proton distributions are approximately uniform 
near the cell edge. In particular, as discussed by Bonche and Vautherin~\cite{bonche1981}, two types of Dirichlet-Neumann boundary conditions are physically 
admissible: either the wavefunction or its radial derivative vanishes at the cell edge. A further simplification is to solve the HF+BCS equations instead of the 
full HFB equations. Systematic calculations have recently shown that the error on the total energy amounts to a few keV per nucleon at most~\cite{pastore2017}.
The Wigner-Seitz approximation allows for relatively fast numerical computations, but is unreliable 
in the densest region of the crust due to the appearance of spurious neutron shell effects~\cite{chamel2007,chamel2008b,margueron2008}. More importantly,  
entrainment between the neutron superfluid and the crust cannot be studied within this approach since nucleons are localized in the Wigner-Seitz cell.
A few three-dimensional HF(+BCS) calculations of the ground-state of cold dense matter have been carried out~\cite{mag02,gog07}, but are still prone to 
spurious shell effects due to the use of a cubic box with strictly periodic boundary conditions (this limitation has been recently analysed in Ref.~\cite{fattoyev2017}). 

\begin{figure}
\begin{center}
\includegraphics[width=0.65\linewidth,keepaspectratio]{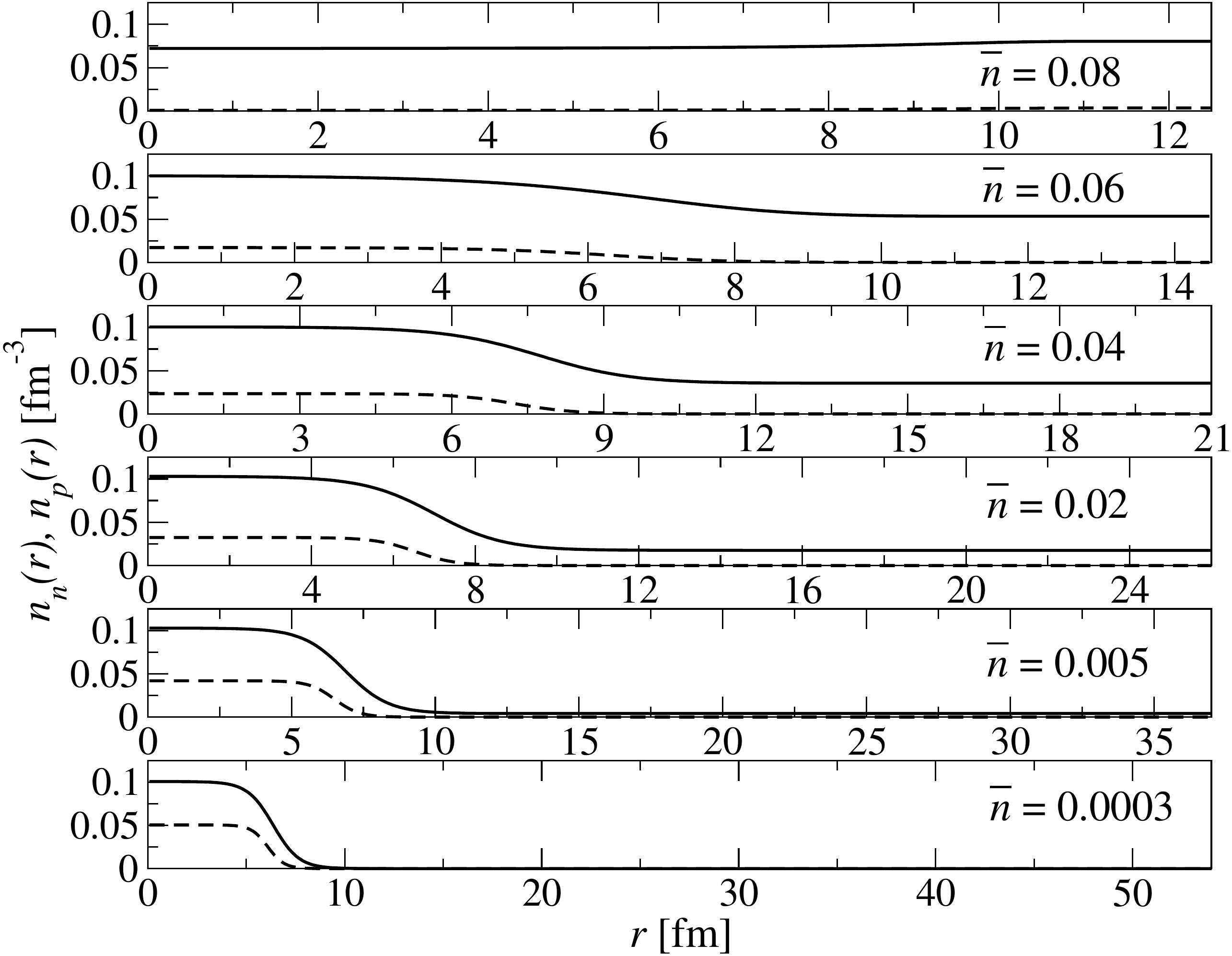}
\end{center}
% \vskip -0.5cm
\caption{Neutron (solid line) and proton (dashed line) density profiles inside the Wigner-Seitz cell for different 
average baryon number densities $\bar n$ (in fm$^{-3}$), as obtained in Ref.~\cite{onsi08}. Note the formation of 
``bubbles'' at $\bar n=0.08$ fm$^{-3}$: the nucleon densities are slightly larger at the cell edge than at the cell center.
Pictures taken from Ref.~\cite{chamel2012}.}
\label{fig1}
\end{figure}

For all these reasons, we have followed a different approach by solving the HF(+BCS) equations perturbatively~\cite{dutta04,onsi08,pea12,pea15}. The main 
contribution to the total energy is determined by the 4th-order Extended Thomas-Fermi (ETF) method (see, e.g., Ref.~\cite{bbd,bch}). 
Namely, the kinetic densities $\tau_q(\pmb{r})$ and the spin-current densities $\pmb{J_q}(\pmb{r})$ are expanded in terms of the nucleon densities 
and their gradients. The total energy $E$ of the system thus reduces to a functional of $n_q(\pmb{r})$, $\pmb{\nabla} n_q(\pmb{r})$ and $\nabla^2 n_q(\pmb{r})$ only, 
treated as the basic variables (instead of the q.p. wavefunctions). 
The minimization of the energy is further simplified by adopting the Wigner-Seitz approximation for the calculation of the Coulomb energy, and by using 
parametrized nucleon density distributions. In particular, we have been employing the following ansatz~\cite{onsi08}
\begin{equation}\label{eq:parametrized-profiles1}
n_q(r) = n_{Bq} + n_{\Lambda q}f_q(r)  \, ,
\end{equation}
in which $n_{Bq}$ is a constant background term, while 
\begin{equation}\label{eq:parametrized-profiles2}
f_q(r) = \frac{1}{1 + \exp \left\{\Big(\frac{C_q - R}
{r - R}\Big)^2 - 1\right\} \exp \Big(\frac{r-C_q}{a_q}\Big) }\, ,
\end{equation}
and $n_{\Lambda q}$, $C_q$ and $a_q$ are free parameters. This form was chosen so as to ensure the vanishing the density gradient at the cell edge. 
The main correction $\delta E$ to the ETF energy arises from proton shell effects. 
Because protons are all bound inside clusters, their Bloch states are essentially independent of $\pmb{k}$. Neutron shell effects are expected 
to be much smaller than proton shell effects (except possibly near the neutron-drip point~\cite{oya94}), and are therefore neglected. 
The correction $\delta E$ is calculated via the Strutinsky integral (SI) theorem (see, e.g., Ref.~\cite{pea15})
\begin{eqnarray}\label{eq:si}
\delta E &=&  \sum\limits_{\alpha}\,V^{(p) 2}_{\alpha} \varepsilon^{(p)}_\alpha
 -\int d^3\pmb{r}\biggl[\overline{B_p}(\pmb{r}) \overline{\tau_p}(\pmb{r}) + \overline{n_p}(\pmb{r})\overline{U_p}(\pmb{r})
+\overline{\pmb{J_p}}(\pmb{r})\cdot \overline{\pmb{W_p}}(\pmb{r})\biggr]\nonumber \\ 
&&-\sum\limits_{\alpha}\,\frac{\Delta^{(p) 2}_\alpha}{4 E^{(p)}_\alpha}  \, , 
\end{eqnarray}
in which overlined fields are the smooth fields emerging from the ETF calculation. 
In this equation, the sums go over the s.p.  states, with the s.p. energies $\varepsilon^{(p)}_\alpha$ being the eigenvalues of the Schr\"odinger equation
\begin{equation}\label{eq:si-eq}
\sum_{\sigma^\prime=\pm1}\overline{h_p}(\pmb{r})_{\sigma\sigma^\prime} \varphi^{(p)}_{\alpha}(\pmb{r},\sigma^\prime)=
\varepsilon^{(p)}_\alpha \varphi^{(p)}_{\alpha}(\pmb{r}) \, ,
\end{equation}
\begin{equation}
\overline{h_p}(\pmb{r})_{\sigma\sigma^\prime} \equiv
-\pmb{\nabla}\overline{B_p}(\pmb{r})\cdot\pmb{\nabla}\delta_{\sigma\sigma^\prime} +
\overline{U_p}(\pmb{r})\delta_{\sigma\sigma^\prime} - {\rm i}\,\overline{\pmb{W_p}}(\pmb{r})\cdot\pmb{\nabla}
\times \pmb{\sigma}_{\sigma\sigma^\prime}\, , 
\end{equation}
while $E^{(p)}_\alpha$ and $V^{(p)}_\alpha$ are the BCS q.p. energies and occupation factors given by Eqs.~(\ref{eq:Uk-Vk}) and (\ref{eq:Eqp}) respectively. 
The proton chemical potential $\mu^p$ and the pairing gaps $\Delta^{(p)}_{\alpha}$ are determined self-consistently by solving the BCS Eqs.~(\ref{eq:Nq-BCS}) and 
(\ref{eq:BCS-gap}). This so-called ETFSI method (extended Thomas-Fermi+Strutinsky integral) is a computationally high-speed approximation to the  
self-consistent HF+BCS equations, thus allowing for systematic calculations of the ground-state structure of the neutron-star crust. Results of such calculations 
presented in Ref.~\cite{onsi08} using the Brussels-Montreal EDF BSk14~\cite{sg07} are summarized in Table~\ref{tab1}. The neutron and proton density distributions 
are shown in Fig.~\ref{fig1} for a few crustal layers. As can be seen in Table~\ref{tab1}, the composition of the nuclear clusters constituting the inner crust of a 
neutron star crucially depends on the underlying proton shell structure. The EDFs employed in calculations of neutron-star crusts should thus be carefully chosen. 
The series of Brussels-Montreal EDFs have been specifically developed for astrophysical applications 
(see, e.g., Ref.~\cite{chamel2015c} for a short review). In particular, the BSk14 EDF was fitted to the measured masses of 2149 nuclei with $N,Z\geq8$ from 
the 2003 Atomic Mass Evaluation~\cite{audi03} with a root mean square deviation of 0.73 MeV (the deviation falling to 0.64 MeV for the subset of 185 neutron-rich nuclei with neutron 
separation energy $S_n\leq 5$~MeV). At the same time, an optimal fit to 782 measured values of charge radii was ensured with a root-mean square deviation of 0.03 fm. 
Moreover, the incompressibility $K_v$ of symmetric nuclear matter at saturation was required to fall in the empirical range $240\pm10$~MeV~\cite{col04}. The values for 
the symmetry energy coefficient at saturation and its slope, which play an important role for determining the structure of neutron-star crusts~\cite{grill2012}, 
are consistent with various constraints inferred from both experiments and astrophysical observations~\cite{lattimer2012}. In addition, this EDF was constrained 
to reproduce the equation of state of neutron matter, as calculated by Friedman and Pandharipande~\cite{fp81} using realistic two- and three-body forces. Incidentally, 
this equation of state is in good agreement with more recent calculations~\cite{apr98,ger10,heb10,tew13} at densities relevant to the neutron-star crusts.

\begin{table}
\begin{center}
\begin{tabular}{|c|c|c|c|c|}
\hline
$\bar n$ (fm$^{-3}$) & $\rho$ (g cm$^{-3}$) &$Z$ &  $A$ & $A_{\rm cell}$ \\
\hline
0.0003   & $4.98\times10^{11}$ & 50 & 170 &   200  \\
0.001    & $1.66\times10^{12}$ & 50 & 179 &   460  \\
0.005    & $8.33\times10^{12}$ & 50 & 198 &  1140   \\
0.01     & $1.66\times10^{13}$ & 40 & 170 &  1215   \\
0.02     & $3.32\times10^{13}$ & 40 & 180 &  1485   \\
0.03     & $4.98\times10^{13}$ & 40 & 173 &  1590   \\
0.04     & $6.66\times10^{13}$ & 40 & 216 &  1610   \\
0.05     & $8.33\times10^{13}$ & 20 & 87  &   800   \\
0.06     & $1.00\times10^{14}$ & 20 & 85  &   780   \\
0.07     & $1.17\times10^{14}$ & 20 & 76  &   714   \\
0.08     & $1.33\times10^{14}$ & 20 & 65  &   665   \\
\hline
\end{tabular}
\caption{Composition of the inner crust of a neutron star as obtained in Ref.~\cite{onsi08}: average baryon number density $\bar n$, average mass density $\rho$, 
proton number $Z$ and nucleon number $A$ in each cluster, total number of nucleons $A_{\rm cell}$ in the Wigner-Seitz cell.} 
\label{tab1}
\end{center}
\end{table}

\begin{table}
\begin{center}
\begin{tabular}{|c|c|c|c|}
\hline
$\bar n$ (fm$^{-3}$) & $T_c$ (K) & $\xi$ (fm) & $\Delta^{(n)}_{\rm F}/\varepsilon^{(n)}_{\rm F}$ \\
\hline
0.0003 & $7.9\times 10^8$    & 12.1 &  0.48 \\
0.001  & $3.7\times 10^9$    & 6.1  &  0.40 \\
0.005  & $8.7\times 10^9$    & 5.0  &  0.26 \\
0.01   & $1.0\times 10^{10}$ & 5.2  &  0.19 \\
0.02   & $1.1\times 10^{10}$ & 6.2  &  0.13 \\
0.03   & $1.1\times 10^{10}$ & 7.5  &  0.09 \\
0.04   & $9.5\times 10^{9}$  & 9.4  &  0.07 \\
0.05   & $7.7\times 10^{9}$  & 12.6 &  0.05 \\
0.06   & $5.5\times 10^{9}$  & 18.6 &  0.03 \\
0.07   & $3.3\times 10^{9}$  & 32.5 &  0.02 \\
0.08   & $3.9\times 10^{8}$  & 304  &  0.002 \\
\hline
\end{tabular}
\caption{Properties of the neutron superfluid in the inner crust of a neutron star ignoring the influence of nuclear clusters. For each average baryon number 
density $\bar n$, are shown the critical temperature $T_c$, the coherence length $\xi$ and the ratio of the pairing gap $\Delta^{(n)}_{\rm F}$ to the Fermi 
energy $\varepsilon^{(n)}_{\rm F}$ using the crustal composition of Ref.~\cite{onsi08}. See text for detail.} 
\label{tab2}
\end{center}
\end{table}

\begin{figure}
\begin{center}
\includegraphics[width=0.65\linewidth,keepaspectratio]{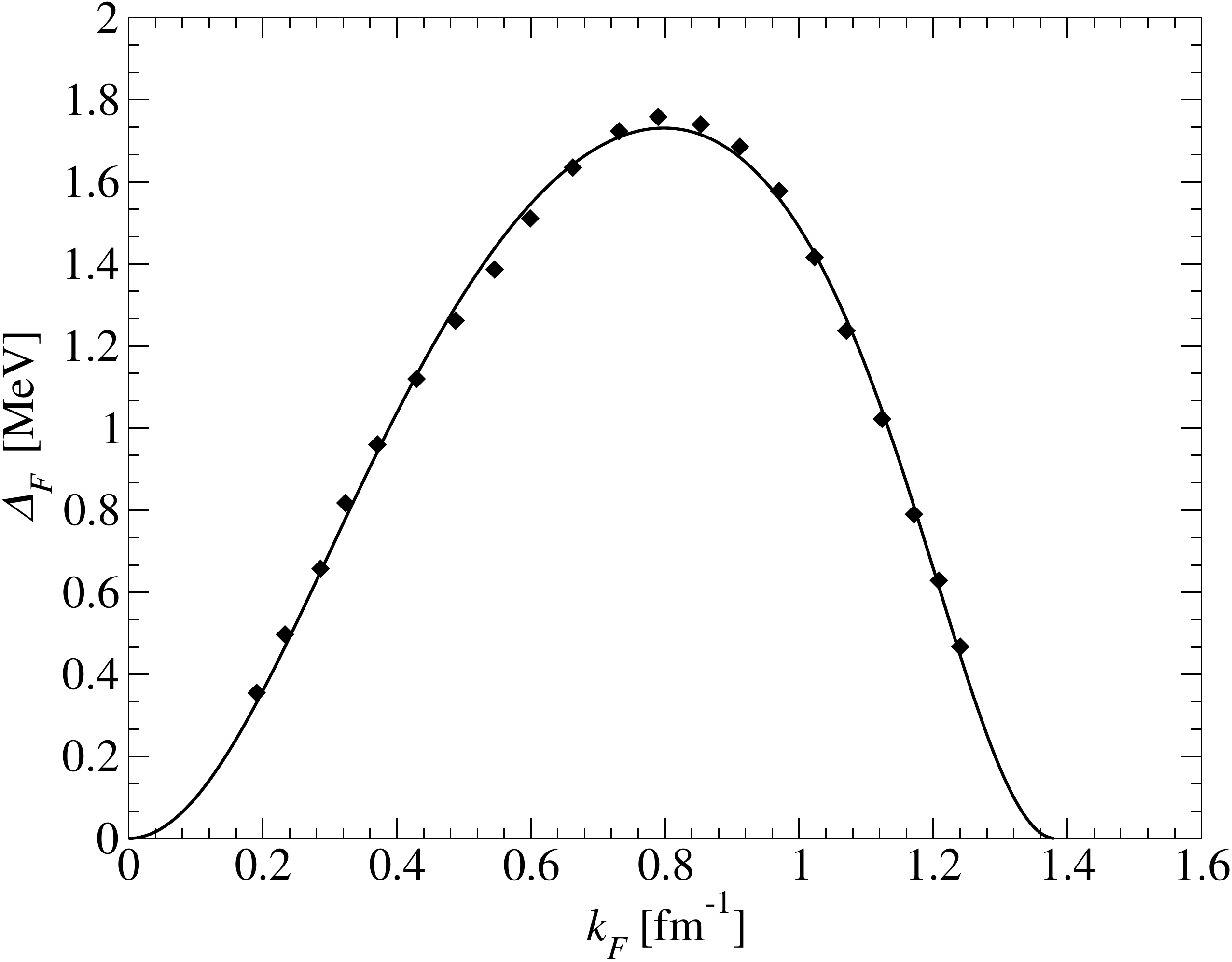}
\end{center}
\caption{$^1$S$_0$ pairing gaps in neutron matter, as obtained by Cao et al. \cite{cao2006} including  
self-energy and medium polarization effects (symbols).  The curve represents a fit to their calculations. }
\label{fig2}
\end{figure}

The superfluid phase transition in uniform neutron matter has been studied using various many-body methods thus providing a benchmark for nuclear EDFs 
(see, e.g., Ref.~\cite{gezerlis2014} for a review). The results of such microscopic calculations have been also widely used to estimate the 
properties of the neutron superfluid permeating the inner crust of a neutron star by neglecting the influence of nuclear clusters and treating 
unbound neutrons as pure neutron matter. Using the crustal composition of Ref.~\cite{onsi08}, and the $^1$S$_0$ neutron pairing gaps $\Delta^{(n)}_{\rm F}$ 
from Ref.~\cite{cao2006} based on the Brueckner theory and shown in Fig.~\ref{fig2}, we have thus calculated at each baryon number density $\bar n$, the critical 
temperature of the neutron superfluid as 
\begin{equation}
T_c(\bar n)=\frac{\exp(\zeta)}{\pi} \Delta^{(n)}_{\rm F}(n_{ n}^{\rm f})\, ,
\end{equation}
with $\zeta\simeq 0.577$ the Euler-Mascheroni constant, and $n_{ n}^{\rm f}=n_{Bn}$ is the density of free neutrons. Similarly, we have calculated the coherence 
length~\cite{bcs57}
\begin{equation}
\xi = \frac{2 \varepsilon^{(n)}_{\rm F}}{\pi k_{\rm F} \Delta^{(n)}_{\rm F}}\, ,
\end{equation}
where 
\begin{equation}
\varepsilon^{(n)}_{\rm F} = \frac{\hbar^2 k_{\rm F}^2}{2 m_n^*}
\end{equation}
is the neutron Fermi energy, $k_{\rm F}=(3\pi^2 n_{ n}^{\rm f})^{1/3}$ the Fermi wavevector, and $m_n^*$ is the microscopic neutron effective mass
(not to be confused with the dynamical effective masses introduced in Section~\ref{sec:hydro-equiv}). The latter was 
obtained from extended Brueckner-Hartree-Fock calculations using the interpolation of Ref.~\cite{lns}. Results are summarized in Table~\ref{tab2}. As can be 
seen in Fig.~\ref{fig1}, the coherence length $\xi$ is of the same order as the size of spatial inhomogeneities, or even larger, especially in the deep regions of 
the crust. For this reason, the presence of nuclear clusters may change substantially the neutron superfluid properties. 
The neutron superfluid transition was first studied within the band theory of solids in Refs.~\cite{chamel2013,chamel2010} by solving the BCS gap 
Eqs.~(\ref{eq:Nq-BCS}) and (\ref{eq:BCS-gap}) for the neutrons. The s.p. states were calculated by solving the Schr\"odinger equation
\begin{equation}\label{eq:neutron-Hamiltonian}
\Big\{-\pmb{\nabla}\overline{B_n}(\pmb{r})\cdot\pmb{\nabla} +
\overline{U_n}(\pmb{r})\Big\} \varphi^{(n)}_{\alpha\pmb{k}}(\pmb{r},\sigma)=
\varepsilon^{(n)}_{\alpha\pmb{k}} \varphi^{(n)}_{\alpha\pmb{k}}(\pmb{r},\sigma) \, ,
\end{equation}
with Bloch boundary conditions
\begin{eqnarray}
 \varphi^{(n)}_{\alpha\pmb{k}}(\pmb{r}+\pmb{\ell}, \sigma)=\exp({\rm i} \pmb{k}\cdot\pmb{\ell})\,\varphi^{(n)}_{\alpha\pmb{k}}(\pmb{r}, \sigma)\, ,
\end{eqnarray}
using the s.p. fields $\overline{B_n}(\pmb{r})$ and $\overline{U_n}(\pmb{r})$ obtained from the ETFSI calculations. The spin-orbit potential 
$\overline{\pmb{W_n}}(\pmb{r})$, which is proportional to $\pmb{\nabla}\overline{n_q}(\pmb{r})$ is small in most region of the inner crust 
(nuclear clusters in the neutron-star crust have a very diffuse surface so that the spin-orbit potential is much smaller than that in isolated nuclei~\cite{chamel2007}), 
and was therefore neglected for simplicity. The crust was assumed to be a perfect body-centered cubic lattice, as in the outer crust. 
Calculations were performed in the dense regions of the crust where the Wigner-Seitz approximation breaks down. The neutron superfluid in 
neutron-star crusts bears similarities with terrestrial multiband superconductors such as magnesium diboride. The main difference lies in the fact 
that the number of bands involved in the pairing phenomenon can be huge (up to $\sim 1000$) due to the strong nuclear attraction. In particular, 
both bound and unbound neutrons are paired, and should thus be treated consistently. Because of the large coherence length as compared to the size of clusters, 
proximity effects are very important. As a result, pairing correlations are substantially enhanced inside clusters while they are reduced in the 
intersticial region, leading to a smooth spatial variation of the pairing potential. The presence of clusters was found to reduce the average neutron 
pairing gap at the Fermi level and the critical temperature by $\sim 20$~\%. The impact of clusters on the superfluid dynamics is much more dramatic. 

Despite the absence of viscous drag, the neutron superfluid flow can still be affected by the crust. These effects were studied in 
Refs.~\cite{epstein1988,sedrakian1996,magierski2004} by calculating the classical potential flow of a neutron liquid induced by the motion of a single cluster.  
For simplicity, the neutron liquid was assumed to be incompressible with density $n_{ n}^{\rm f}$ and the cluster was treated as a uniform density 
sphere of radius $R_{\rm I}$. Except in Ref.~\cite{sedrakian1996}, the cluster was supposed to be permeable to the neutron liquid, an hypothesis consistent 
with microscopic calculations. With these approximations, the hydrodynamical equations can be analytically solved. The neutron superfluid density is 
expressible as
\begin{equation}\label{eq:n_n^s-hydro}
 n_{ n}^{\rm S}=n_{ n} - N^* n_{\rm I}\, ,
\end{equation}
where $n_{\rm I}$ is the cluster number density, and the effective number of neutrons $N^*$ in a cluster is given by 
\begin{equation}
N^*= N \frac{(1-\gamma)^2}{1+2\gamma}\, , 
\end{equation}
with $N=(4/3) \pi R_{\rm I}^3 n_{ n}^{\rm I}$ the number of neutrons in a cluster whose neutron density is $n_{ n}^{\rm I}$, and 
$\gamma=n_{ n}^{\rm f}/n_{ n}^{\rm I}$. In this model, the average neutron number density is given by 
\begin{equation}
 n_{ n}=\frac{\mathcal{V}_{\rm I}}{\mathcal{V}_{\rm cell}} n_{ n}^{\rm I} + \left(1-\frac{\mathcal{V}_{\rm I}}{\mathcal{V}_{\rm cell}}\right) n_{ n}^{\rm f}\, ,
\end{equation}
where $\mathcal{V}_{\rm I}=(4/3) \pi R_{\rm I}^3$, and $\mathcal{V}_{\rm cell}=1/n_{\rm I}$ is the volume of the Wigner-Seitz cell. 
The neutron superfluid density can be equivalently expressed as 
\begin{equation}\label{eq:n_n^s-hydro2}
\frac{n_{ n}^{\rm S}}{n_{ n}^{\rm f}}=1+3\frac{\mathcal{V}_{\rm I}}{\mathcal{V}_{\rm cell}}\frac{1-\gamma}{1+2\gamma} \, .
\end{equation}
Leaving aside the possibility of nuclear bubble, we have $\gamma\leq 1$ so that $N^*\leq N$: the neutrons in the cluster move with an effectively 
reduced speed due to the counterflow of liquid through the cluster. In this simple model, the neutron superfluid is not entrained by the crust but 
counter moves. The neutron superfluid density is thus found to be larger than the density of free neutrons
\begin{equation}
1\leq \frac{n_{ n}^{\rm S}}{n_{ n}^{\rm f}}\leq 1+3\frac{\mathcal{V}_{\rm I}}{\mathcal{V}_{\rm cell}} \, .
\end{equation}
However, these results should be interpreted with some care. Indeed, as shown in Ref.~\cite{martin2016}, the neutron number $N$ does not generally 
coincide with the number of neutrons that are actually bound in the cluster in the quantum mechanical sense (i.e. a state is quantum mechanically bound if 
its s.p. energy $\varepsilon^{(n)}_{\alpha\pmb{k}}$ lies below the maximum of the potential $U_n(\pmb{r})$). The number $N$ was actually found to systematically 
overestimate the number of quantum mechanically bound neutrons, by up to a factor $\sim 3.5$ at average baryon number density $\bar n=0.06$~fm$^{-3}$. 
The neutron flow induced by a periodic lattice of clusters has been recently studied in Ref.~\cite{martin2016} under the same assumptions as in 
Refs.~\cite{epstein1988,sedrakian1996,magierski2004}. The resulting neutron superfluid density is essentially the same as that given by 
Eq.~(\ref{eq:n_n^s-hydro2}). This conclusion was actually anticipated in Ref.~\cite{epstein1988} given that the lattice spacing is typically 
much larger than $R_{\rm I}$. In this analysis, the density $n_{ n}^{\rm I}$ was defined as the physical density of neutrons located in the cluster. 
However, as first pointed out in Ref.~\cite{epstein1988}, this density should rather be interpreted as a neutron superfluid density in the cluster, 
which may be different from $n_{ n}^{\rm I}$. Introducing the fraction $\delta$ of superfluid neutrons in the cluster, the effective number of 
neutrons in the cluster becomes~\cite{martin2016}
\begin{equation}
N^*= N \left(1-\delta + \frac{(\delta-\gamma)^2}{\delta+2\gamma}\right)\, .
\end{equation}
Inserting this expression in Eq.~(\ref{eq:n_n^s-hydro}) yields 
\begin{equation}\label{eq:n_n^s-hydro3}
\frac{n_{ n}^{\rm S}}{n_{ n}^{\rm f}}=1+3\frac{\mathcal{V}_{\rm I}}{\mathcal{V}_{\rm cell}}\frac{\delta-\gamma}{\delta+2\gamma} \, .
\end{equation}
The superfluid density is \emph{smaller} than the density of free neutrons if $\delta < \gamma$, in which case the neutron superfluid is entrained by 
the crust. Allowing the neutrons in the cluster to bo partially superfluid extends the range of $n_{ n}^{\rm S}/n_{ n}^{\rm f}$: 
\begin{equation}
1-\frac{3}{2}\frac{\mathcal{V}_{\rm I}}{\mathcal{V}_{\rm cell}}\leq \frac{n_{ n}^{\rm S}}{n_{ n}^{\rm f}}\leq 1+3\frac{\mathcal{V}_{\rm I}}{\mathcal{V}_{\rm cell}} \, .
\end{equation}
The lower bound coincides with the case $\delta=0$ originally considered in Ref.~\cite{sedrakian1996} whereby the cluster is treated as a solid obstacle. 
In this limit, the ratio $n_{ n}^{\rm S}/n_{ n}^{\rm f}$ is independent of $\gamma$. In all regions of the crust but the deepest, 
$\mathcal{V}_{\rm I}\ll\mathcal{V}_{\rm cell}$ (see, e.g. Fig.~\ref{fig1}), so that the hydrodynamical models predict 
$n_{ n}^{\rm S} \sim n_{ n}^{\rm f}$. 

The local hydrodynamical approximation assumes that the neutron superfluid coherence length $\xi$ is much smaller than $R_{\rm I}$, 
a condition that is however usually not fulfilled in any region of the inner crust, as can be seen from Fig.~\ref{fig1} and Table~\ref{tab2} 
(see, also Ref.~\cite{martin2016}). The first quantum 
mechanical calculations of entrainment effects were presented in Refs.~\cite{chamel2005,chamel2006b,carter2005c} using the band theory of solids. 
Treating the crust as a polycrystalline solid and averaging over all directions, the neutron superfluid density is given by~\cite{carter2005d} 
\begin{equation}\label{eq:super-density}
n_{ n}^{\rm S}=\frac{m}{24\pi^3\hbar^2}\sum_\alpha \int{\rm d}^3k\, |\pmb{\nabla}_{\pmb{k}}
\varepsilon^{(n)}_{\alpha\pmb{k}}|^2 \frac{(\Delta^{(n)}_{\alpha\pmb{k}})^2}{(E^{(n)}_{\alpha\pmb{k}})^3}\, ,
\end{equation}
where the integral is taken over the first Brillouin zone, and $\pmb{\nabla}_{\pmb{k}}$ denotes the gradient in $\pmb{k}$-space. 
In the weak coupling limit $\Delta^{(n)}_{\alpha\pmb{k}}\ll\varepsilon^{(n)}_{\rm F}$, 
the neutron superfluid density reduces to an integral over the neutron Fermi surface (defined by the set of $\pmb{k}$ points 
such that $\varepsilon^{(n)}_{\alpha\pmb{k}}=\mu^{ n}$)~\cite{carter2005c} 
\begin{eqnarray}\label{eq:super-density-weak}
n_{ n}^{\rm S} &\approx& \frac{m}{12\pi^3\hbar^2}\sum_\alpha \int{\rm d}^3k\, |\pmb{\nabla}_{\pmb{k}}\varepsilon^{(n)}_{\alpha\pmb{k}}|^2 
\delta(\varepsilon^{(n)}_{\alpha\pmb{k}}-\mu^{ n})\nonumber \\ &=& \frac{m}{12\pi^3\hbar^2}\sum_\alpha \int_{\rm F} |\pmb{\nabla}_{\pmb{k}} \, 
\varepsilon^{(n)}_{\alpha\pmb{k}}|{\rm d}{\cal S}^{(\alpha)} \, .
\end{eqnarray}
The neutron superfluid density can be equivalently expressed as the trace of an effective mass tensor similar to that originally 
introduced in solid-state physics for electrons (see, e.g., Ref.~\cite{ash76}) 
\begin{equation}\label{eq:effmass-tensor}
 \left(\frac{1}{m_n^*(\pmb{k})^\alpha}\right)_{ij} = \frac{1}{\hbar^2}\frac{\partial^2 \varepsilon^{(n)}_{\alpha\pmb{k}}}{\partial k_i\partial k_j}\, ,
\end{equation}
\begin{equation}
n_{ n}^{\rm S}=\frac{1}{12\pi^3}\sum_\alpha \int_{\rm F} {\rm d}^3k\, {\rm Tr}\biggl[\frac{m}{m_n^*(\pmb{k})^\alpha}\biggr]\, ,
\end{equation}
where the integral is taken over the Fermi volume (defined by the set of $\pmb{k}$ points such that $\varepsilon^{(n)}_{\alpha\pmb{k}}\leq \mu^{ n}$). 
The concept of effective mass tensor~(\ref{eq:effmass-tensor}) has been also employed in the context of neutron diffraction in ordinary 
crystals~\cite{ze86,ra95}. 
Entrainment effects can be alternatively formulated in terms of the effective number $A^\star$ of nucleons attached to clusters~\cite{chamel2013}
\begin{equation}
 A^\star = A_{\rm cell} - \frac{n_{ n}^{\rm S}}{n_{\rm I}}\, .
\end{equation}

\begin{figure}
\begin{center}
\includegraphics[width=0.9\linewidth,keepaspectratio]{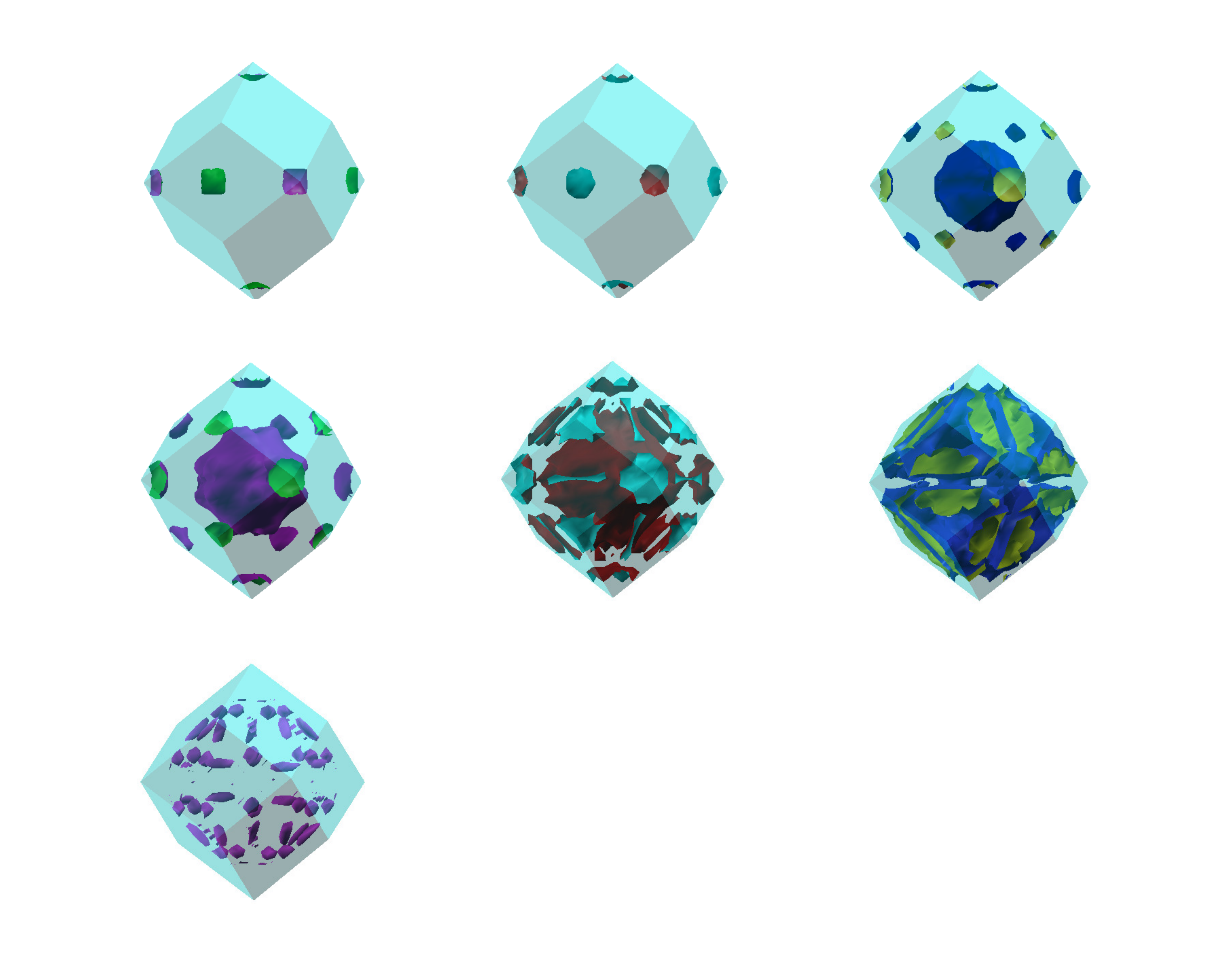}
\end{center}
% \vskip -0.5cm
\caption{Neutron Fermi surface in the crust of a neutron star at average baryon number density $\bar n=0.0003$ fm$^{-3}$ in the reduced zone 
scheme: each panel shows a different branch of the Fermi surface in the first Brillouin zone. 
Figure made with XCrySDen~\cite{xcrysden} using the neutron band structure calculated in Ref.~\cite{chamel2012}.}
\label{fig3}
\end{figure}

\begin{figure}
\begin{center}
\includegraphics[width=0.9\linewidth,keepaspectratio]{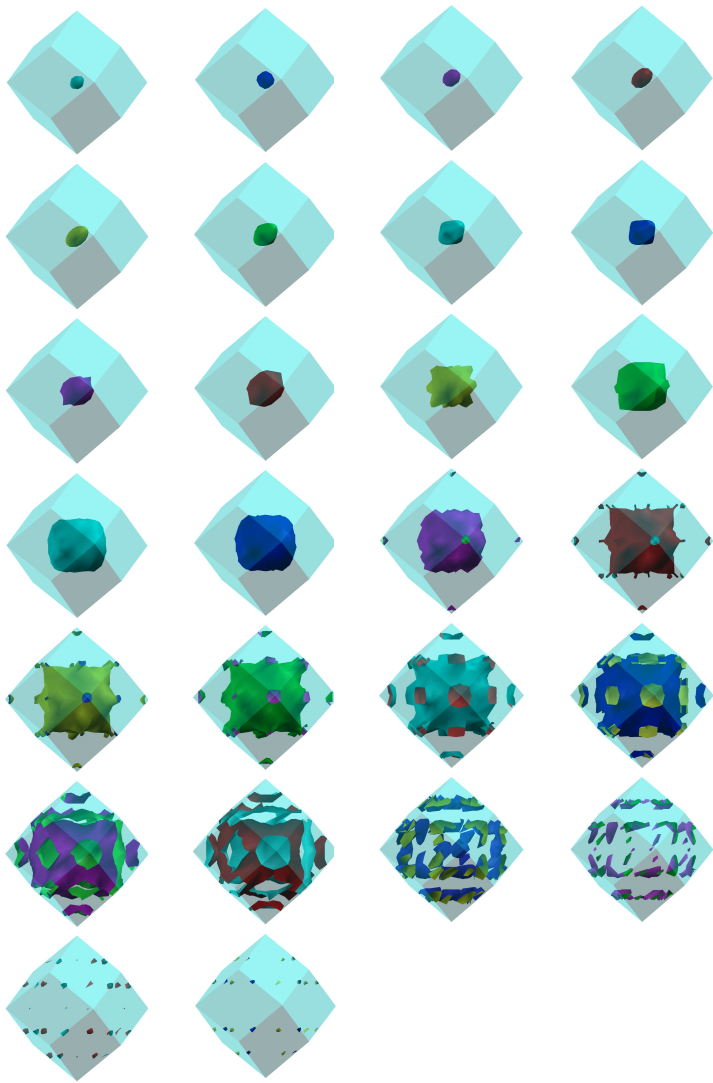}
\end{center}
% \vskip -0.5cm
\caption{Same as Fig.~\ref{fig3} at average baryon number density $\bar n=0.03$ fm$^{-3}$.
Figure made with XCrySDen~\cite{xcrysden} using the neutron band structure calculated in 
Ref.~\cite{chamel2012}.}
\label{fig4}
\end{figure}

Systematic band-structure calculations in all regions of the inner crust of a neutron star using the crustal composition previously 
obtained in Ref.~\cite{onsi08} were carried out (see Ref.~\cite{chamel2012} for numerical detail). Results are summarized in Table~\ref{tab3}. In all 
regions of the crust, the neutron superfluid density is found to be lower than the density of unbound neutrons: the neutron superfluid is therefore entrained by the crust. 
Similarly to the case of electrons in ordinary solids, the transport properties of free neutrons are governed by the shape of the neutron Fermi 
surface, which in turn depends on the lattice interactions (unlike the Fermi volume given by $\mathcal{V}_{\rm F}=(2\pi)^3 n_{ n}^{\rm f}$). In 
the shallowest layer in the vicinity of the neutron drip transition, the neutron Fermi wavelength $\lambda_{\rm F}=2\pi/k_{\rm F}$ is much larger that 
the lattice spacing so that the Fermi volume is entirely contained inside the first Brillouin zone, and the Fermi surface is nearly spherical. The neutron 
superfluid can thus flow freely through the crust and $n_{ n}^{\rm S}\sim n_{ n}^{\rm f}$. 
With further compression, the neutron Fermi volume increases until it touches the Brillouin zone boundary. For a body-centered cubic lattice, this occurs as the 
density of unbound neutrons reaches the threshold value $n_{ n}^{\rm f}=n_{\rm I} \sqrt{2}\pi/3$ (about 1.5 unbound neutrons per lattice site). At this 
point, the Fermi surface is expected to be substantially distorted by the periodic potential recalling that a wave vector $\pmb{k}$ lying on a zone boundary 
satisfies the diffraction condition $2 \pmb{k}\cdot\pmb{G} = G^2$ where $\pmb{G}$ denotes a reciprocal lattice vector: a neutron with wave vector $\pmb{k}$ 
will thus be Bragg-reflected by the lattice. As a consequence, the Fermi surface area is reduced, as shown in Table~\ref{tab3} (a more detailed analysis can 
be found in Refs.~\cite{chamel2006b,chamel2007}). On the contrary, the density of s.p. states at the Fermi level given by 
\begin{equation}
 \mathcal{N}_{\rm F} = \mathcal{V}_{\rm cell}\sum_\alpha \int \frac{{\rm d}^3 k}{(2\pi)^3}\delta(\varepsilon_{\alpha\pmb{k}} - \mu^n) 
 = \mathcal{V}_{\rm cell}\sum_\alpha \int_{\rm F} \frac{{\rm d}{\cal S}^{(\alpha)}}{|\pmb{\nabla}_{\pmb{k}} \, \varepsilon^{(n)}_{\alpha\pmb{k}}|}\, , 
\end{equation}
remains essentially unaffected by the lattice~\cite{chamel2006b,chamel2007}, as can be seen in Table~\ref{tab3}. This quantity is of particular interest 
for determining thermal properties such as the neutron specific heat~\cite{chamel2009}.  
Since the Fermi surface area 
$\mathcal{S}_{\rm F}$ is reduced compared to the corresponding Fermi sphere area $\mathcal{S}^{\rm f}_{\rm F}$, 
the average Fermi velocity $(1/\hbar)|\pmb{\nabla}_{\pmb{k}} \, \varepsilon^{(n)}_{\alpha\pmb{k}}|$ must be reduced by the same amount. From 
Eq.~(\ref{eq:super-density-weak}), we can infer that $n_{ n}^{\rm S}~\sim (\mathcal{S}_{\rm F}/\mathcal{S}^{\rm f}_{\rm F})^2 n_{ n}^{\rm f}$, 
as first pointed out in Ref.~\cite{chamel2005} (see also Ref.~\cite{chamel2009}). 
This scaling is approximately satisfied, as can be seen in Table~\ref{tab3}. From these general considerations, we therefore expect 
$n_{ n}^{\rm S} \leq n_{ n}^{\rm f}$ at variance with results obtained within the local hydrodynamical approximation discussed previously. 
Examples of neutron Fermi surfaces are plotted in Figs.~\ref{fig3} and \ref{fig4} for two different average baryon number densities. Note that the 
Fermi surface has as many different branches as bands satisfying the defining equation $\varepsilon^{(n)}_{\alpha\pmb{k}}=\mu^n$. The distortions of the 
neutron Fermi surface, and in particular the formation of necks, can be clearly seen. 
The more the Fermi surface intersects Brillouin zone boundaries, the larger will generally be the effect of the lattice on the neutron superfluid density. 
The number of intersections depends on the ratio between the Fermi volume and the volume $\mathcal{V}_{\rm BZ}=(2\pi)^3/\mathcal{V}_{\rm cell}$ of the 
first Brillouin zone. Their ratio $\mathcal{V}_{\rm F}/\mathcal{V}_{\rm BZ}=n_{ n}^{\rm f}/n_{\rm I}$ is simply equal to the average 
number of unbound neutrons per lattice site. Basically, this number is the lowest at the neutron drip point, peaks at about $A_{\rm cell}-A=1417$ at density 
$\bar n=0.03$ fm$^{-3}$ and decreases at higher densities. As expected, the neutron superfluid density follows a similar behavior (see Table~\ref{tab3}). 
The same trend has been independently predicted in the context of superfluid atomic gases in optical lattices~\cite{watanabe2008}. 
With increasing density, the lattice interactions become progressively weaker, as can be inferred from Fig.~\ref{fig1}, thus further reducing entrainment 
effects. 

The strong reduction of the neutron superfluid density in the intermediate crustal regions at densities $\bar n\sim 0.03$ fm$^{-3}$ has been 
recently questioned in Refs.~\cite{martin2016,pethick2017} in view of the neglect of neutron pairing in Eq.~(\ref{eq:super-density-weak}). In particular, 
the authors of Ref.~\cite{pethick2017} have solved the HFB Eqs.~(\ref{eq:HFB}) for neutrons in a fixed external periodic potential, and found that to a large extent 
band-structure effects are suppressed by pairing. As a result, the superfluid density is much less reduced than predicted in Ref.~\cite{chamel2012}. This 
conclusion however is puzzling. Indeed, at the densities $\bar n\sim 0.03$ fm$^{-3}$ where entrainment effects are the strongest, the neutron pairing gaps 
are expected to be relatively small $\Delta^{(n)}_{\rm F}/\varepsilon^{(n)}_{\rm F}\sim 10\%$, as can be seen in Table~\ref{tab2}. It therefore seems unlikely that 
calculting the superfluid density using Eq.~(\ref{eq:super-density}) instead of (\ref{eq:super-density-weak}) would lead to dramatically different 
results since the factor $(\Delta^{(n)}_{\alpha\pmb{k}})^2/(E^{(n)}_{\alpha\pmb{k}})^3$ is expected to be strongly peaked at the Fermi surface. On the other hand, the 
calculations of Ref.~\cite{pethick2017} were performed using a simplified model of the neutron-star crust. In particular, the solid crust was approximated by 
a one-dimensional periodic lattice, and the potential $\overline{U_n}(\pmb{r})$, which ressembles a smooth square well around clusters (see Fig.~\ref{fig1}), 
was replaced by a pure sinusoidal potential of the form $\overline{U_n}(z)\approx 2\breve{U}_n(G)\cos(G z)$, where $\breve{U}_n(G)$ is the Fourier coefficient 
of the original potential associated with the reciprocal lattice vector $\pmb{G}$. Although many Fourier components of the original potential are small, keeping 
only one and ignoring all the others may introduce considerable errors. For instance, at the density $\bar n=0.03$ fm$^{-3}$ considered in Ref.~\cite{pethick2017}, 
the depth of the original potential $\overline{U_n}$ is about $\sim 30$ MeV, whereas its individual Fourier components $|\breve{U}_n(G)|\lesssim 2$~MeV. In other 
words, the periodic potential adopted in Ref.~\cite{pethick2017} is an order of magnitude weaker than that originally used in Ref.~\cite{chamel2012}. Moreover, 
the field $\overline{B_n}(\pmb{r})$ appearing in the s.p. Hamiltonian~(\ref{eq:neutron-Hamiltonian}) was replaced by $\hbar^2/(2 m_n)$. Finally, the neutron 
superfluid density was estimated assuming that each Fourier component of the potential contributes independently, and by integrating over $G$ treated as a 
continuous variable. In view of the many approximations, it seems premature to draw general conclusions on the role of pairing. The suppression of band-structure 
effects found in Ref.~\cite{pethick2017} still needs to be confirmed by solving the fully three-dimensional HFB Eqs.~(\ref{eq:HFB}) with Bloch boundary conditions 
using the same Hamiltonian~(\ref{eq:neutron-Hamiltonian}) as that employed in Ref.~\cite{chamel2012}. Quantum and thermal fluctuations of clusters may also 
influence band-structure effects, as suggested in Ref.~\cite{kobyakov2013}. On the other hand, nuclear clusters are effectively heavier due to entrainment 
($A^\star>A$) thus reducing the frequency of lattice vibrations, as shown in Section~\ref{sec:hydro-excitations}. The role of low-energy excitations on 
entrainment needs to be investigated self-consistently.

\begin{table}
\begin{center}
\begin{tabular}{|c|c|c|c|c|}
\hline
$\bar n$ (fm$^{-3}$) & $A^\star$ & $n_{ n}^{\rm S}/n_{ n}^{\rm f}$ (\%) & $\mathcal{S}_{\rm F}/\mathcal{S}^{\rm f}_{\rm F}$ (\%) & $\mathcal{N}_{\rm F}/\mathcal{N}_{\rm F}^{\rm f}$ (\%) \\
\hline
0.0003 &  175  & 82.6 & 92.1 & 107   \\
0.001  &  383  & 27.3 & 49.2 & 104   \\
0.005  &  975  & 17.5 & 38.2 & 99.4  \\
0.01   & 1053  & 15.5 & 36.2 & 100   \\
0.02   & 1389  & 7.37 & 24.3 & 98.9  \\
0.03   & 1486  & 7.33 & 24.6 & 98.1  \\
0.04   & 1462  & 10.6 & 29.9 & 101   \\
0.05   &  586  & 30.0 & 51.5 & 98.6  \\
0.06   &  461  & 45.9 & 63.3 & 96.7  \\
0.07   &  302  & 64.6 & 75.3 & 93.5  \\
0.08   &  247  & 64.8 & 74.3 & 91.9  \\
\hline
\end{tabular}
\caption{Properties of the inner crust of a neutron star as determined by band-structure calculations~\cite{chamel2012}: 
average baryon number density $\bar n$, effective number $A^\star$ of nucleons attached to clusters, 
ratio of the neutron superfluid density $n_{ n}^{\rm S}$ to the density $n_{ n}^{\rm f}$ of unbound neutrons, ratio of the Fermi surface 
area $\mathcal{S}_{\rm F}$ to the area of the Fermi sphere of an ideal neutron Fermi gas with density $n_{ n}^{\rm f}$, ratio of the density of states 
$\mathcal{N}_{\rm F}$ to that of an ideal neutron Fermi gas.} 
\label{tab3}
\end{center}
\end{table}

\section{Conclusion}

Despite the absence of viscous drag, the neutron superfluid present in the inner crust of a neutron star cannot flow freely. 
The neutron superfluid is coupled to the solid crust by nondissipative entrainment effects, whereby the momentum of the neutron 
superfluid is generally not aligned with the neutron velocity similarly to laboratory superfluid atomic gases in optical lattices~\cite{watanabe2008}. Applying the fully 4-dimensionally covariant variational formalism developed in Refs.~\cite{carter2004,carter2005}, we have shown how to construct a minimal smooth-averaged hydrodynamical model 
of superfluid neutron-star crusts, taking into account the effects of entrainment. The equivalence of this formulation with the 
more heuristic approach of Refs.~\cite{pethick2010,kobyakov2013} has been demonstrated. The different treatments of 
entrainment in terms of an entrainment matrix, dynamical effective masses or superfluid density have been clarified. Entrainment may have a profound influence on the superfluid dynamics. For example, we have shown that the Bogoliubov-Anderson excitations of 
the neutron superfluid are strongly mixed with longitudinal lattice vibrations thus illustrating the need for a consistent treatment 
of neutron-star crusts. Entrainment effects have implications for observed astrophysical phenomena, such as pulsar frequency 
glitches.  

A smooth-averaged hydrodynamical description of neutron-star crusts requires the specification of some microscopic inputs, such 
as the static internal energy density $U_{\rm ins}(n_n,n_p)$ and the neutron superfluid density $n_n^{\rm S}$ in the simple model 
presented in this paper. We have shown how to determine these ingredients using the nuclear EDF theory. This 
theory provides a self-consistent quantum description of superfluid neutrons and nuclear clusters, but its full implementation in neutron-star 
crusts remains challenging. For this reason, we have developed a computationally very fast approach~\cite{pea15}, in which the quantum shell effects are treated as a small correction to the total energy. This method allows for systematic calculations of the internal structure of 
neutron-star crusts. For this purpose, we have employed the accurately calibrated Brussels-Montreal EDFs. We have studied the neutron superfluid transition in the framework of the BCS theory of multiband superconductors. 
Because of the strong long-range attractive nuclear interaction, both bound and unbound neutrons form Cooper pairs involving 
up to $\sim 1000$ bands. As a consequence, the pairing mechanism is highly nonlocal. The presence of the nuclear inhomogeneities
reduces the average neutron pairing gap on the Fermi surface $\Delta^{(n)}_{\rm F}$ and the critical temperature $T_c$ by $\sim20$~\%~\cite{chamel2010}. On the 
other hand, the neutron superfluid dynamics is found to be strongly influenced by the nuclear lattice~\cite{chamel2012}.
Systematic band-structure calculations have shown that the neutron superfluid density $n_n^{\rm S}$ is reduced by about an order of magnitude as compared to the density $n_n^{\rm f}$ of unbound neutrons in the intermediate
region of the inner crust at densities $\sim 0.02-0.03$~fm$^{-3}$ so that the neutron superfluid is strongly entrained by the crust. 
These calculations were carried out in the limit $\Delta^{(n)}_{\rm F}/\varepsilon^{(n)}_{\rm F}\rightarrow 0$. Although this approximation
appears reasonable in view of the BCS expression of $n_n^{\rm S}$, Eq.~(\ref{eq:super-density}), and the fact that 
$\Delta^{(n)}_{\rm F}/\varepsilon^{(n)}_{\rm F}\sim 0.1$, the neglect of pairing has been recently questioned~\cite{martin2016,pethick2017}. 
In particular, band-structure effects were found to be suppressed by pairing in Ref.~\cite{pethick2017} considering however a simplified 
model of the crust, whereby neutrons were assumed to interact with a very weak one-dimensional sinusoidal potential. These results need to be confirmed 
with fully three-dimensional calculations using the same realistic periodic potentials as in Ref.~\cite{chamel2012}. More importantly, the 
role of lattice vibrations, impurities, and defects deserve further studies. 

The advantage of the fully covariant formulation developed in Refs.~\cite{carter2004,carter2005} is that it facilitates the comparison with 
the relativistic theory, which will be ultimately required for a realistic global description of neutron stars. Besides, this variational formalism considerably simplifies the derivation 
of conservation laws (e.g. conservation of helicity currents) and identities (e.g. generalised Bernouilli constants and virial theorems) 
making use of differential geometric concepts such as Killing vectors. Dissipative processes can be naturally incorporated along the lines of 
Ref.~\cite{carter2005b}. More importantly, this formalism can be easily extended so as to account for the rigidity of the solid crust, and 
the presence of a strong magnetic field, both within the Newtonian theory~\cite{carter2006,carter2006b} and in the fully relativistic 
context~\cite{carter2006c}. It should be stressed that this formalism is very general and thus could  also be applied to study the dynamics of various laboratory (super)fluid systems.

\begin{acknowledgements}
This work was supported by the Fonds de la Recherche Scientifique - FNRS (Belgium) under grant n$^\circ$~CDR J.0187.16, 
and the European Cooperation in Science and Technology (COST) action MP1304 \emph{NewCompStar}. 
\end{acknowledgements}

\pagebreak

\end{document}